\documentclass{article}

\usepackage{lipsum} 

\usepackage{amsmath}
\usepackage{bm}
\usepackage{multirow}
\usepackage{textcomp}
\usepackage{color}

\usepackage{graphicx}
\usepackage[justification=justified,footnotesize,]{caption}
\usepackage{subcaption}
\usepackage{natbib}
\usepackage{multicol}
\usepackage{setspace}

\usepackage[sc]{mathpazo} 
\usepackage[T1]{fontenc} 
\usepackage{microtype} 

\usepackage[left=.5in,right=.5in,top=.5in,bottom=.5in]{geometry}
\usepackage{multicol} 
\usepackage{booktabs} 
\usepackage{float} 

\usepackage{lettrine} 
\usepackage{paralist} 

\usepackage{abstract} 

\usepackage{titlesec} 
\renewcommand\thesection{\arabic{section}} 
\renewcommand\thesubsection{\arabic{section}.\arabic{subsection}} 
\titleformat{\section}[block]{\large\bfseries\scshape}{\thesection.}{1em}{}
\titleformat{\subsection}[block]{\large\slshape}{\thesubsection.}{1em}{}
\titlespacing{\section}{0em}{1em}{.5em}

\newcommand{\ER}{Erd\H{o}s-R\'{e}nyi }

\title{\vspace{-10mm}\fontsize{20pt}{15pt}\selectfont\textbf{ Empirical Reference Distributions for \\[2.5mm]
 Networks of Different Size\\}} 

\author{
\large
\textsc{Anna Smith}$^{\text{a},}$\textsc{, Catherine A. Calder}$^{\text{a}}$\footnote{Corresponding author:  calder@stat.osu.edu} \textsc{, Christopher R. Browning}$^{\text{b}}$\\[2mm] 
\normalsize $^{\text{a}}$\textit{Department of Statistics, The Ohio State University} \\ 
\normalsize $^{\text{b}}$\textit{Department of Sociology, The Ohio State University} \\
\vspace{-8mm}
}

\begin{document}

\maketitle 


\vspace{-5mm}
\begin{abstract}
\vspace{-.05in}
Network analysis has become an increasingly prevalent research tool across a vast range of scientific fields.  Here, we focus on the particular issue of comparing network statistics, i.e. graph-level measures of network structural features, across multiple networks that differ in size.  Although ``normalized'' versions of some network statistics exist, we demonstrate via simulation why direct comparison is often inappropriate.  We consider normalizing network statistics relative to a simple fully parameterized reference distribution and demonstrate via simulation how this is an improvement over direct comparison, but still sometimes problematic.  We propose a new adjustment method based on a reference distribution constructed as a mixture model of random graphs which reflect the dependence structure exhibited in the observed networks.  We show that using simple Bernoulli models as mixture components in this reference distribution can provide adjusted network statistics that are relatively comparable across different network sizes but still describe interesting features of networks, and that this can be accomplished at relatively low computational expense.  Finally, we apply this methodology to a collection of ecological networks derived from the Los Angeles Family and Neighborhood Survey activity location data.\\

\end{abstract}

\vskip4ex




\section{Introduction}

\begin{figure*}[t]
\caption{Realizations from a Single Model}
\centering
\includegraphics[width=.3\textwidth]{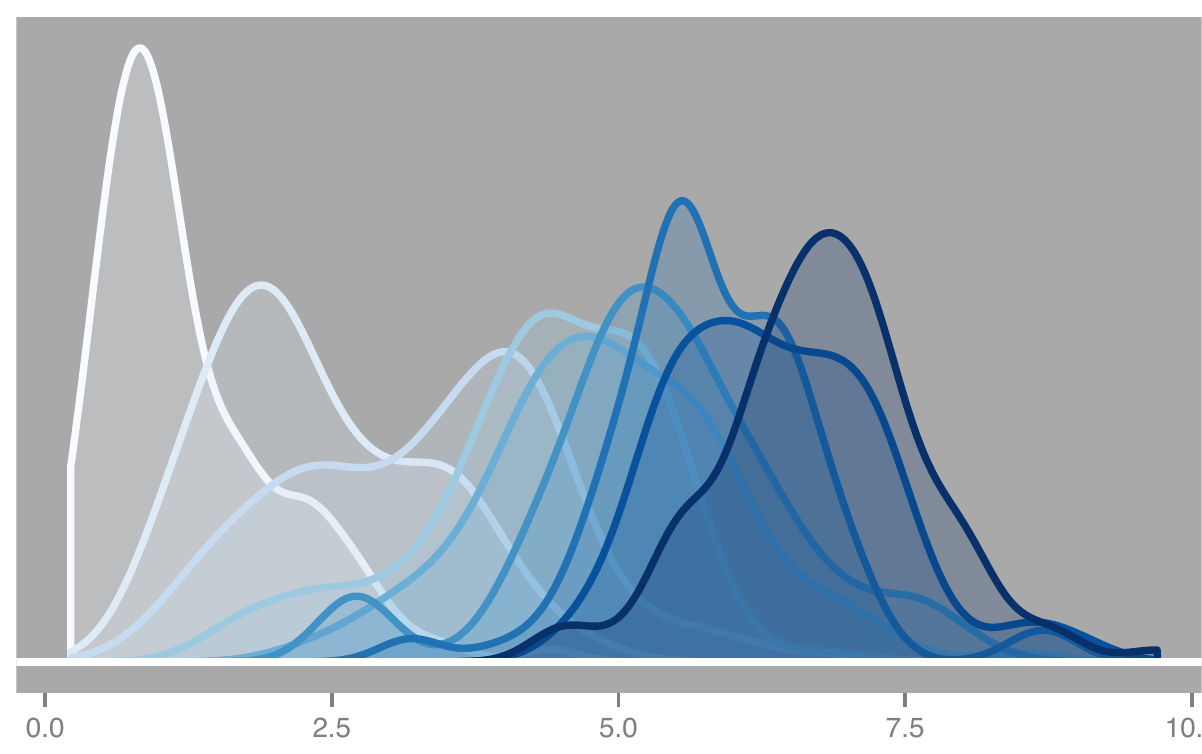}
\\
\includegraphics[width=.35\textwidth]{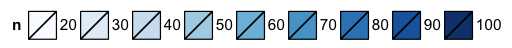}
\vspace{1mm}
\caption*{The distribution of a network statistic (here, closeness centrality) across a range of network sizes (number of nodes, $n$).}
\label{fig:intro}
\end{figure*}

Networks have become ubiqutous as a tool for analysis in many areas of applied research.  Here, we focus on the situation where the researcher is interested in considering and comparing multiple networks.  Often the networks under consideration consist of differing numbers of nodes, and the comparison of statistics across these networks is not straightforward.  That is, the distribution of a network statistic across networks may be very different, depending on the size of the network.  This holds even when the networks are generated from a single model, as shown in Figure \ref{fig:intro}.

A tool for network comparisons would be useful in many fields - from traditional social network analysis to brain network studies in biomedical research.  For example, many researchers have compared friendship networks within different classrooms in an effort to infer patterns of friendship formation and their role in academic performance \citep{lubbers_2003, lubbers_snijders_2007}.  Similarly, a growing area of study in biomedical research attempts to draw inference about patterns of disease, illness, and treatment effectiveness from brain networks by comparing networks across affected and unaffected patients \citep[e.g.][]{bassett_bullmore_verchinski_etal_2008, he_chen_evans_2008}.  Of course, many similar potential applications exist:  comparing e-mail communication networks among employees in different departments of a company, comparing citation networks among researchers in different fields, comparing voting networks among politicians from different state congresses, etc.

Further, it would be advantageous to have measures of network structural features that can be included as covariates in typical regression models.  For example, it might be interesting to relate the average shortest path length of an individual's brain network to his/her stage of Alzheimer's disease, or to relate the degree of transitivity (proportion of times the co-votee of my co-votee also co-votes with me) in a state congress' voting network to the number of bipartisan legislative acts passed in that state.  In order to include such network measures in a regression-style analysis, the measures must be on the same scale and have the same meaning across the various networks.  However, when the networks being compared vary in size, often the values of these network statistics can be dominated by size effects.

We first demonstrate via simulation why direct comparison of network statistics across networks of different sizes is often inappropriate.  We then investigate the use of an absolute reference distribution which is fully parameterized a priori, here choosing to compare to completely random graphs, or \ER graphs, of the appropriate size \citep[e.g.][]{vanwijk_stam_daffertshofer_2010, stam_2004, schindler_bialonski_horstmann_etal_2008, smit_stam_posthuma_etal_2008}.  We demonstrate via simulation why this is an improvement over direct comparison, but still sometimes problematic.  We argue that a more appropriate method of comparison would both adjust for network size as well as incorporate some degree of appropriate network dependence structure.  We propose utilizing a new reference distribution which comes from a mixture model, where we are mixing over the dependence structure demonstrated in a random subset of the empirically observed networks.  To accomplish this, we consider three models for reference distribution simulations:  the simple Bernoulli model, a mean degree preserving Bernoulli model \citep{krivitsky_handcock_morris_2011}, and a Hierarchical Bernoulli model, where the latter is a special case of the HERGM, or Hierarchical ERGM, proposed by \citet{schweinberger_handcock_2015}.  We consider the performance of these methods in a series of simulation studies.  Finally, we apply these techniques to neighborhood network data from the Los Angeles Family and Neighborhood Survey (L.A.FANS) to demonstrate the type of substantive statements that these methods can facilitate.


\section{Background and Motivation}

We consider networks or graphs, $G$, that consist of $n$ nodes or vertices and represent them by their $n \times n$ adjacency matrix, $\bm{Y}$.  Each element of this matrix, $Y_{ij}$, is a tie indicator variable such that
\[ Y_{ij} = \left\{ \begin{array}{ll}
1 & \text{if nodes $i$ and $j$ are tied} \\
0 & \text{otherwise} \end{array} \right. \hspace{10mm} \text{for $i,j=1,...n$} \]
For simplicity, we consider only one-mode, undirected and unweighted networks, and so we let $Y_{ii} = 0$ by definition.  Certainly, the methodology proposed here could be extended to other types of networks.  We refer to the number of nodes, $n$, as network size, but note that others denote this property as the \textit{order} of the graph \citep[e.g][]{kolaczyk_2014}.

Generally, the issue of network comparisons has recieved little methodological attention, despite the popularity of network analysis.  Further, much of researchers' attention has been focused on building methods to model multiple networks simultaneously and using this modeling framework as a way to proceed with any questions of network comparison, similar to a traditional meta-analysis approach \citep{snijders_baerveldt_2003, an_2015}.  This style of analysis requires faith in the individual micro-level analyses themselves.  However, working with existing network models can sometimes be difficult since these models (even in the simplest cases):  often suffer from degeneracy \citep[for some discussion, see][]{snijders_pattison_robins_2006, chatterjee_diaconis_2013}, can have dependence structures that are difficult to interpret \citep[such as the ERGM terms suggested by][]{snijders_pattison_robins_2006}, face computational issues, and, most importantly, can exhibit a rather striking lack of connection between model parameters and the specific structural features of networks generated from these models.  In particular, many common network model parameters are often highly correlated so that individual parameters play different roles under different contexts.  Consider the example highlighted by \cite{snijders_pattison_robins_2006} of an ERGM with parameters for only edges and triangles: ``for a fixed positive transitivity parameter, as the edge parameter becomes more negative there is a point at which... [simulated networks] change dramatically... from only complete graphs to only low density graphs''.  In this simple ERGM, the edges parameter does not seem to affect network density in the way that we might think, or at least the strength and smoothness of its effect appears to depend on the particular model specification itself.  Thus, even in simple versions of existing network models, there are cases where model parameters do not appear to characterize the structural aspects of networks which they seem designed to describe.  For this reason, when comparing multiple networks, we propose comparing direct measures of network structure itself - network or graph statistics - as a simple alternative to comparing fitted model parameter values whose interpretations are often not straight forward.  The methodology we propose here could also be valuable as an additional viewpoint in combination with the existing meta-analysis style technique \citep[e.g. as decribed by][]{snijders_baerveldt_2003, an_2015}, where our proposed methodology could perhaps guide the specification of micro-level analyses and help to improve overall interpretability.


\subsection{Common Network Statistics} \label{sec:stats}

We consider network statistics that are popular in the applied network science literature \citep{anderson_butts_carley_1999, smith_moody_2013}.  The statistics we consider here can be thought of as describing two different aspects of a network:  centralization and topology.  We consider degree, closeness, and betweenness as measures of centralization while our considered measures of topology are average path length and transitivity.  We also consider network density.  

All statistics are computed (and compared) at the graph level.  For the centralization measures, we use Freeman's formula \citep{freeman_1979}:
\[  C(G) = \sum_{j=1}^n \left\{ \max_{1 \le i \le n} s(i) - s(j) \right\} \]
for any vertex-level statistic, $s(i)$.  Note that at the graph level, $C(G)$ is a measure of the dispersion of the distribution of $s(i)$, across all vertices in the graph.  We consider this ``unnormalized'' version of our centralization measures as well as the ``normalized'' version in which $C(G)$ is divided by the maximum possible value attainable by $C(G)$ for a graph of the same size.  Unfortunately, the normalized version proposed by \cite{butts_2006} which adjusts for both the size and density of $G$ has not yet been implemented in standard software to our knowledge and so it will not be considered here.

\textbf{Degree Centrality} at the vertex level is simply a node's degree, or the number of other nodes to which that node is tied:
\[ s_{\text{deg}} (i) = \sum_{j \ne i} Y_{ij}. \]
Thus, at the vertex level, this statistic reports the size of the $i$th node's personal or immediate network so that at the graph level, we measure the dispersion in personal network size.

\textbf{Closeness Centrality} at the vertex level is the reciprocal of the sum of the shortest path lengths from that node to all other nodes:
\[ s_{\text{clo}} (i) = \frac{1}{ \sum_{j \ne i} d_{ij} } \]
where $d_{ij}$ is the length of the shortest path from node $i$ to node $j$.  Intuitively, $s_{clo}(i)$ captures the speed or efficiency with which information, disease, etc. could be spread from vertex $i$ to the rest of the nodes in $G$.  At the graph level, we measure the amount that this efficiency varies across all nodes.

\textbf{Betweenness Centrality} at the vertex level is the proportion of shortest paths between all pairs of nodes that pass through the node of interest:
\[ s_{\text{btw}} (i) = \sum_{k \ne j; \>\> k,j \ne i} \frac{ g_{kij} }{ \tilde{g}_{kj} } \]
where $g_{kij}$ is the number of paths from node $k$ to node $j$ that pass through node $i$, while $\tilde{g}_{kj}$ counts all paths from $k$ to $j$.  Adding an additional degree of complexity to closeness centrality, betweenness centrality at the vertex level measures the importance of the $i$th node as a central node in efficient communication or spread across $G$.  Again, at the graph level, we get the degree of variation in this measure of a node's importance.

\textbf{Average Path Length} is a graph-level measure and is defined as the average of the shortest path lengths between all pairs of nodes:
\[ P(G) = \frac{1}{n-1} \sum_{i=1}^n\sum_{j \ne i} d_{ij}.  \]
Similar to closeness centrality, we again are interested in the speed or efficiency with which nodes are able to transmit information across $G$.  Recall that at the graph level, closeness centrality measures the dispersion of these efficiencies.  Average path length, however, instead captures the ``average efficiency'', where graphs with larger $P(G)$ are considered less efficient than graphs with smaller $P(G)$. 

\begin{figure}
\caption{Subgraphs included in Transitivity Network Statistic}
\centering
\begin{subfigure}{.075\textwidth}
\includegraphics[width=.95\linewidth]{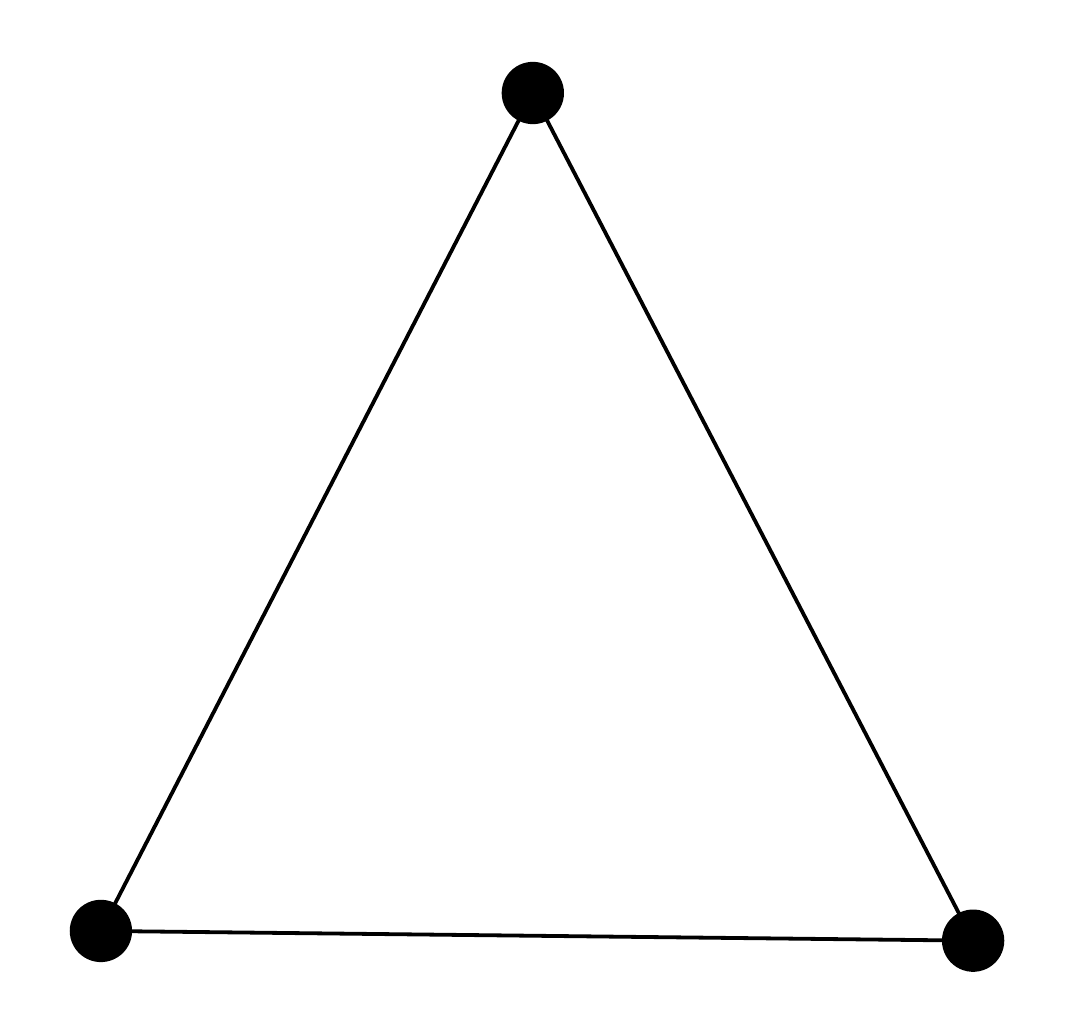}
\caption*{Triangle}
\end{subfigure}
\hspace{5mm}
\begin{subfigure}{.075\textwidth}
\includegraphics[width=.95\linewidth]{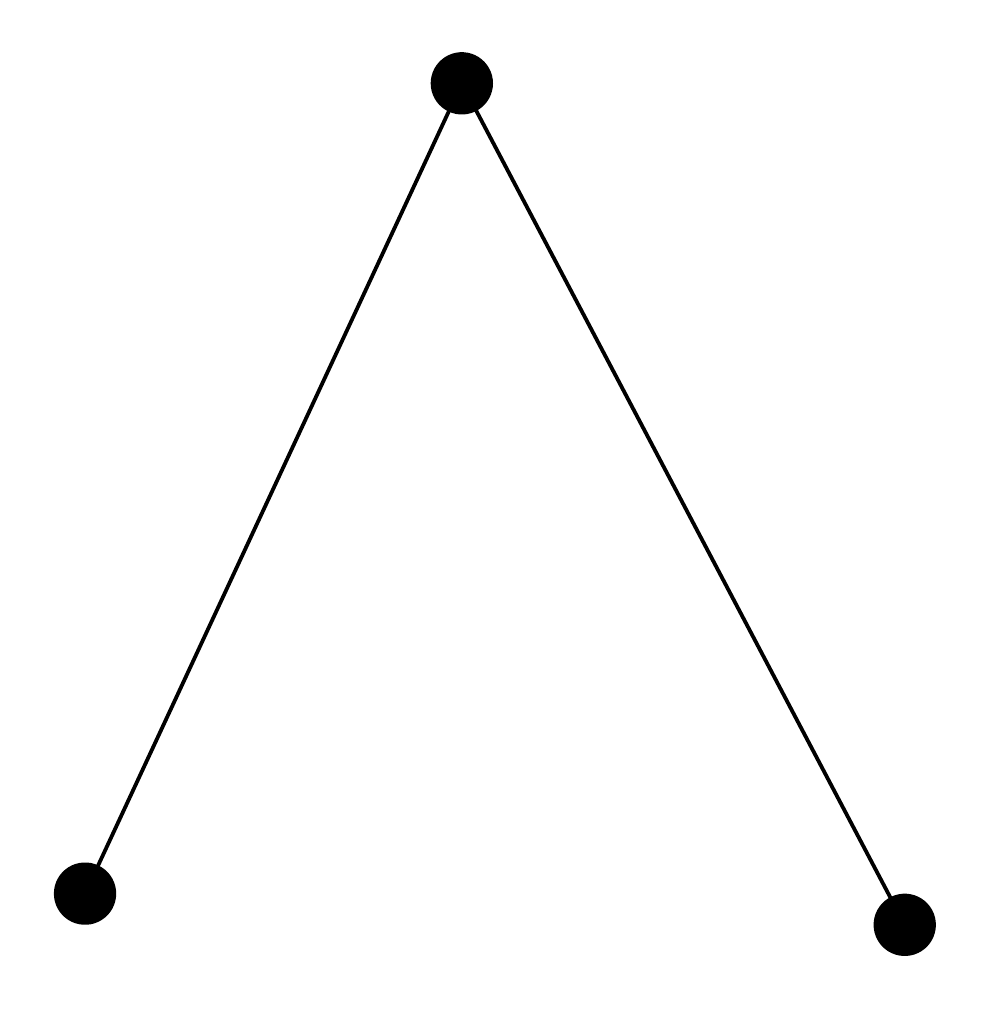}
\caption*{2-star}
\end{subfigure}
\label{fig:Stars}
\end{figure}

\textbf{Transitivity}, also called Clustering, at the vertex level is the proportion of 2-stars centered at node $i$ which are closed triangles.  At the graph level, we simply report the average:
\[ T(G) = \frac{ \text{\# of closed triangles}}{ \text{\# of 2-stars}} \]
where triangles and 2-stars are the subgraph types depicted in Figure \ref{fig:Stars}.  This statistic is extremely popular in applied social network research, largely due to its relevance to much of social theory.  Intuitively, $T(G)$ reflects the average amount of local clustering in $G$.

\textbf{Density} is a graph-level measure of the proportion of all possible ties that are realized in the observed network:
\[ D(G) = \frac{1}{ {n \choose 2} } \sum_{i < j } Y_{ij}. \]
Sometimes treated as a fixed attribute, similarly to the number of nodes, we instead prefer to treat density in the same manner as all other network statistics examined here.  Note that other work sometimes uses the terms density and mean degree interchangeably \citep[e.g][]{anderson_butts_carley_1999, vanwijk_stam_daffertshofer_2010}, where mean degree is $\frac{1}{n} \sum_{i=1}^n \sum_{j\ne i} Y_{ij} = (n-1) \> D(G)$. 

Despite the popularity of these common network statistics, there has been very little research that addresses their dependence on network size.  \citet{anderson_butts_carley_1999} examine the distributions of a variety of graph-level network statistics (including degree and betweenness centrality) across a range of network sizes.  However, they consider only networks generated from one particular model:  the conditional uniform graph (CUG), where they allow both the network size and mean degree to vary.  Generally, a CUG model is a uniform distribution across all possible graphs that share some common feature, such as degree distribution, density, etc.  Using a CUG model conditional on shared network size and mean degree, \citeauthor{anderson_butts_carley_1999} provide evidence that the distributions of these network statistics clearly depend on network size, and that they are also influenced by network density.  Continuing in this line of work, we will examine distributions of these common network statistics across a range of generative network models, including some that better mimic real world network data than conditional uniform graphs.

\citet{faust_2006} examines a wide range of real world networks (across a variety of applications and settings) to show that the triad census can be largely described by lower order graph properties, including size and density.  The triad census is an example of a subgraph census, a commonly used measure of local graph structure.  More specifically, the triad census is the collection of counts for each possible configuration of three (unlabeled) nodes across all triads in the graph of interest.  \citeauthor{faust_2006}'s result that the triad census can be explained by lower order graph properties implies that other, higher-order network properties (perhaps such as the common network statistics mentioned here) might also be heavily influenced by network size.

Here, we will be primarily building off of the work of \citet{vanwijk_stam_daffertshofer_2010}.  The authors provide a detailed summary of the problems faced in comparing network statistics across multiple networks as well as a brief summary of a range of possible solutions, all through the lens of neuroscience research.  They focus solely on the topological measures, average path length and transitivity, as those capture network features important in brain network theory.  In fact, the authors provide closed form adjustments for these two statistics for networks simulated from particular network models \citep[Table 1]{vanwijk_stam_daffertshofer_2010}.  However, as \citeauthor{vanwijk_stam_daffertshofer_2010} point out, the underlying generative model is typically unknown for empirical, real world networks.

Besides these few works, the issue of comparing network statistics across networks has recieved little methodological attention.  Perhaps this is the result of simply ignoring a lesser-known issue.  In fact, many of the network statistics examined here \textit{appear} to be ``normalized'' in some way that incorporates network size, so that it is certainly plausible that a reasearcher might assume that these statistics ought to be comparable across sizes.  In the case of the centralization measures, ``normalized'' versions are acheived by dividing by the largest possible value attainable on a graph \textit{of that size}.  The topological measures are versions of averages and are thus at least roughly adjusted for size effects.  However, as we will see in the following section, these simple adjustments are not enough to account for what appears to be a complicated dependence between these common network statistics and network size.


\subsection{Generative Models for Networks} \label{sec:models}

One might hope that statistics computed on networks that come from the same generative model would have comparable values, regardless of network size.  However, as has been shown for relatively simple classes of generative models \citep{anderson_butts_carley_1999, vanwijk_stam_daffertshofer_2010} and as we will see for a range of other models, this is certainly not the case.  To examine this issue, we simulate groups of networks that cover a range of sizes \textit{from the same model} and examine the distribution of the common network statistics across the range of network sizes.

In an effort to capture the variety in structure, density, topology, etc. which is demonstrated in real world networks, we consider six network models of increasing complexity:  the \ER model, a Bernoulli model, a mean degree preserving Bernoulli model \citep{krivitsky_handcock_morris_2011}, a Markov ERGM, a Hierarchical Bernoulli model and a Hierarchical Markov ERGM.  For simplicitly, we will represent each model as a probability distribution for a random adjacency matrix, $\bm{Y}$.  Alternatively, note that this gives us a joint distribution for the set of possible edges,  $\left\{ Y_{ij} : i < j; \>i,j = 1,...n \right\}$.  Since we are dealing exclusively with undirected graphs, we need only model $\frac{1}{2}$ of the ${n \choose 2}$ possible edge variables, given that $Y_{ji}=Y_{ij} \> \> \forall i, j$ by definition in undirected networks.

Before describing these generative models in detail, we briefly outline how they will be used in the simulation studies in Section \ref{sec:dircomp}, where we monitor the performance of directly comparing statistics across networks of different size.  For each generative model, we simulate $N_n = 200$ replicates of graphs with $n=20, 30, ... ,100$ nodes, so that each model gives rise to a total of $N=1800$ graphs.  This approach of using simulated network data allows us much greater levels of replication within network sizes than would working with any collection of real world networks.

\textbf{\ER Model}. This model is included as a simple baseline.  Although there are a few formulations \citep{erdos_renyi_1960, gilbert_1959}, we focus on the version that is equivalent to a Bernoulli graph with the probability of a tie, $p$, set to $0.50$.  In a Bernoulli graph, each tie indicator variable is modeled as an independent, identically distributed Bernoulli random variable, with $P(Y_{ij}=1)=p$.  This implies that
\begin{align}
P( \bm{Y} = \bm{y} | \theta_{edge} ) = \text{exp} \left\{ \theta_{edge} \> h_{edge} \left( \bm{y} \right) - \psi(\theta_{edge}) \right\}  
\end{align}
where $\theta_{edge} = \text{log} \left( \frac{p}{1-p} \right)$, $h_{edge}(\bm{y})$ counts the number of edges in the observed network, and $\psi(\theta_{edge})$ is the normalizing term, which is calculated as follows:
\[ \psi(\theta_{edge}) = \sum_{\bm{y}^* \in \mathcal{Y}} \text{exp} \left\{ \theta_{edge} \>h_{edge}(\bm{y}^*) \right\} \]
where $\mathcal{Y}$ is the set of all possible graphs of size $n$.  Although much is known about the properties and behavior of this model, it does not tend to reproduce the type of structure or topological features that we typicaly observe in real world networks.

 \textbf{Bernoulli Model}.  We can generalize the \ER model by changing the tie density, so that it better reflects what we often see in real world networks.  Here, we set the probability of a tie, $p$, to be 0.20.  Although we obtain reasonable density under this model, we would still not expect to see the type of topological structure that we associate with real world networks.  This follows from the Bernoulli model's (unrealistic) assumption that all tie variables are independent.

\textbf{Mean Degree Preserving Bernoulli Model}.  We implement the offset terms suggested by \citet{krivitsky_handcock_morris_2011}, which results in an reparameterized version of the traditional Bernoulli model:
\begin{align}
P( \bm{Y} = \bm{y} | \theta_{deg} ) = \text{exp} \left\{ \theta_{deg} \> h_{edge} \left( \bm{y} \right) + \text{log} \left( \frac{1}{n} \right)h_{edge} \left( \bm{y} \right) - \psi(\theta_{deg}) \right\}  
\end{align}
where $h_{edge}(\bm{y})$ counts the number of edges in the observed network as before, and $\psi(\theta_{deg})$ is the normalizing constant:
\[ \psi(\theta_{deg}) = \sum_{\bm{y}^* \in \mathcal{Y}} \text{exp} \left\{ \theta_{deg} \>h_{edge}(\bm{y}^*) + \text{log} \left( \frac{1}{n} \right) h_{edge} \left( \bm{y}^* \right) \right\} \]
The authors show that without an offset term, a Bernoulli model preserves density as network size increases and argue that this is not an appropriate effect for many real-world social networks (e.g. we would not expect the observed average number of friendships per person to increase at the same rate as the sample size).  Instead, models that include the proposed offset are designed to maintain mean degree\footnote{Note that mean degree scales linearly with density by a factor of $n-1$} as $n$ increases.  \cite{krivitsky_handcock_morris_2011} show that under this model, the degree distribution converges to a Poisson distribution as $n \rightarrow \infty$, with average degree exp($\theta_{deg}$).  We choose $\theta_{deg}$ such that the average degree is three.

\begin{figure*}[t]
\caption{Realizations from Simulated Datasets, $n=20$}
\centering
\begin{subfigure}{.15\textwidth}
\includegraphics[width=.95\linewidth]{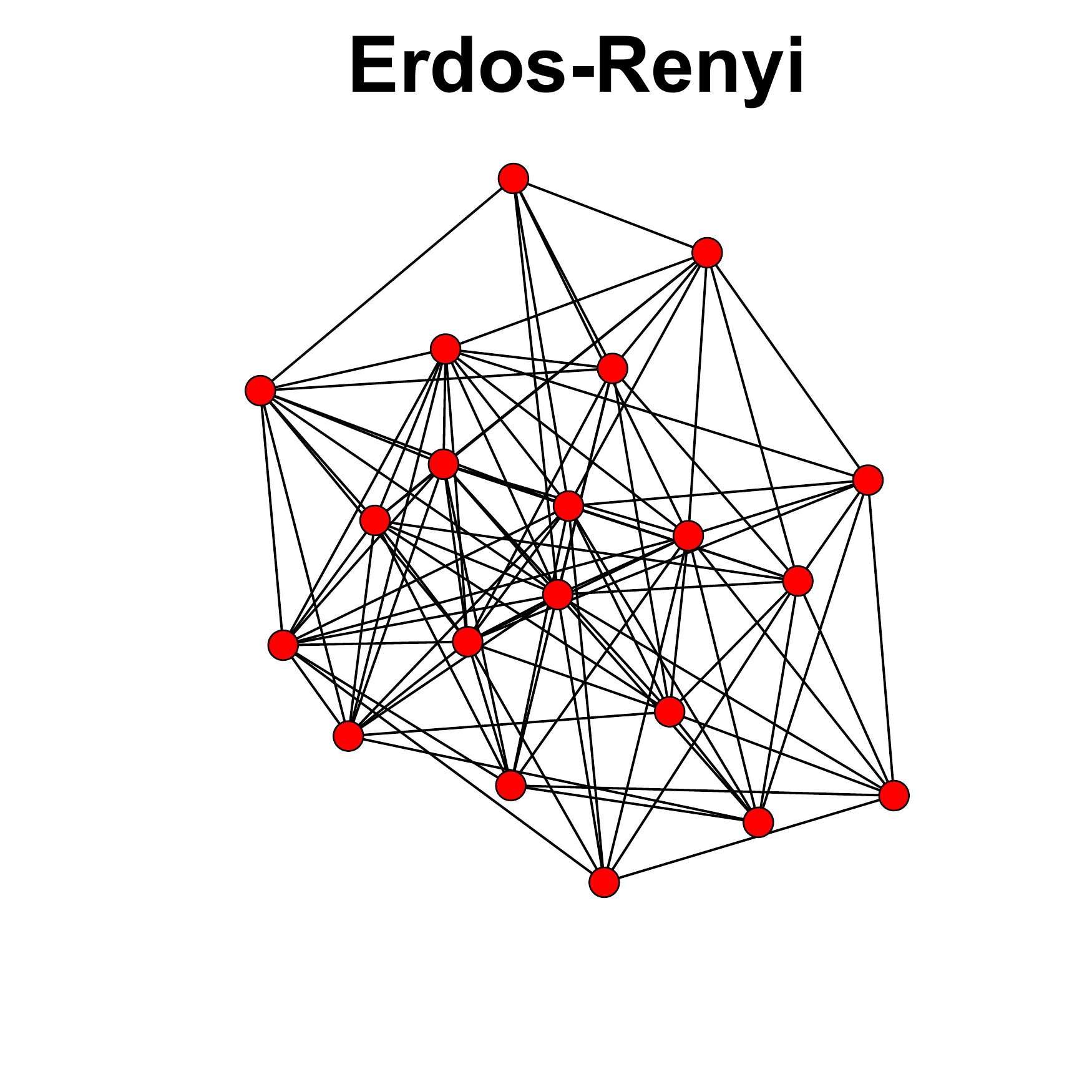}
\caption*{\scriptsize$p=0.50$}
\end{subfigure}
\begin{subfigure}{.15\textwidth}
\includegraphics[width=.95\linewidth]{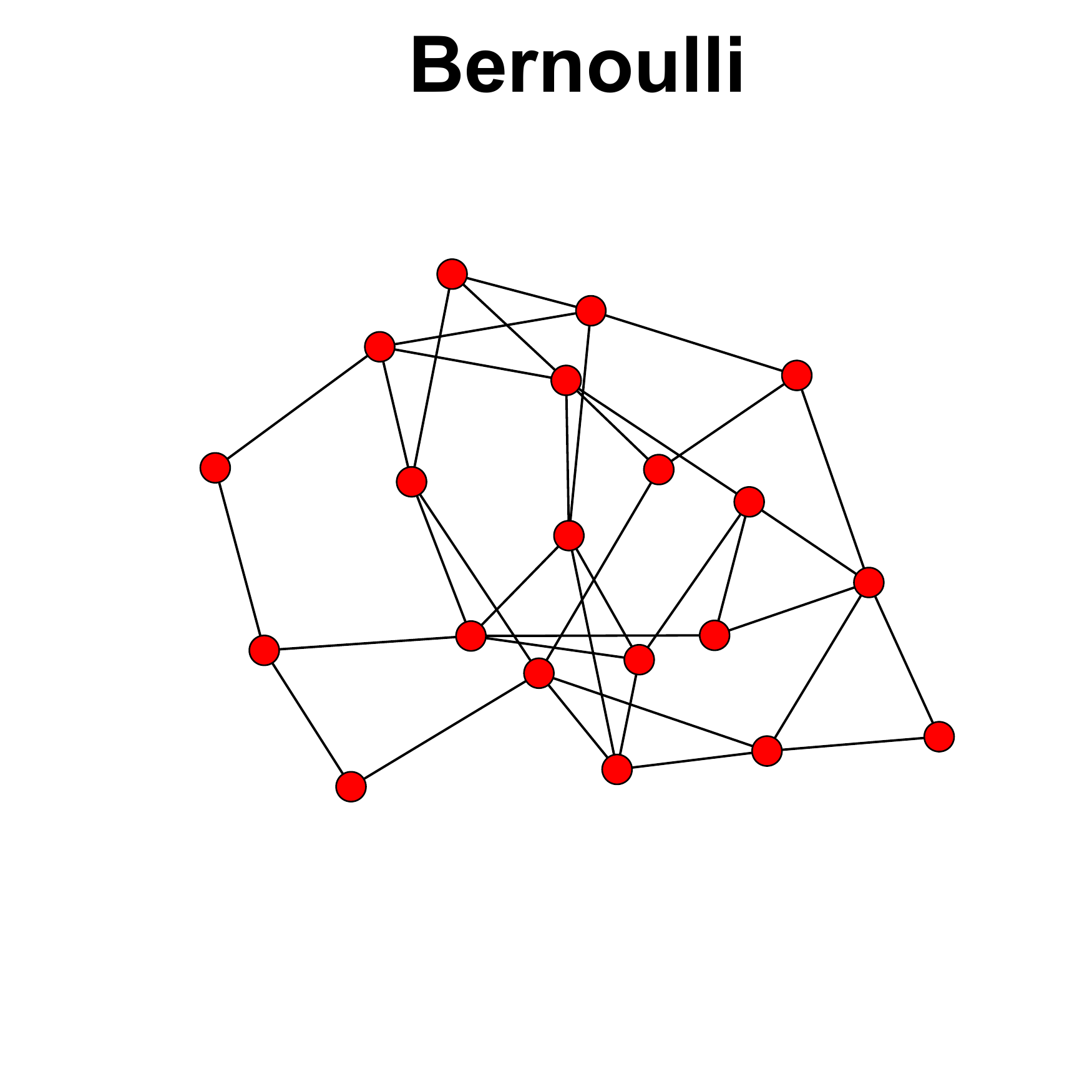}
\caption*{\scriptsize$p=0.20$}
\end{subfigure}
\begin{subfigure}{.15\textwidth}
\includegraphics[width=.95\linewidth]{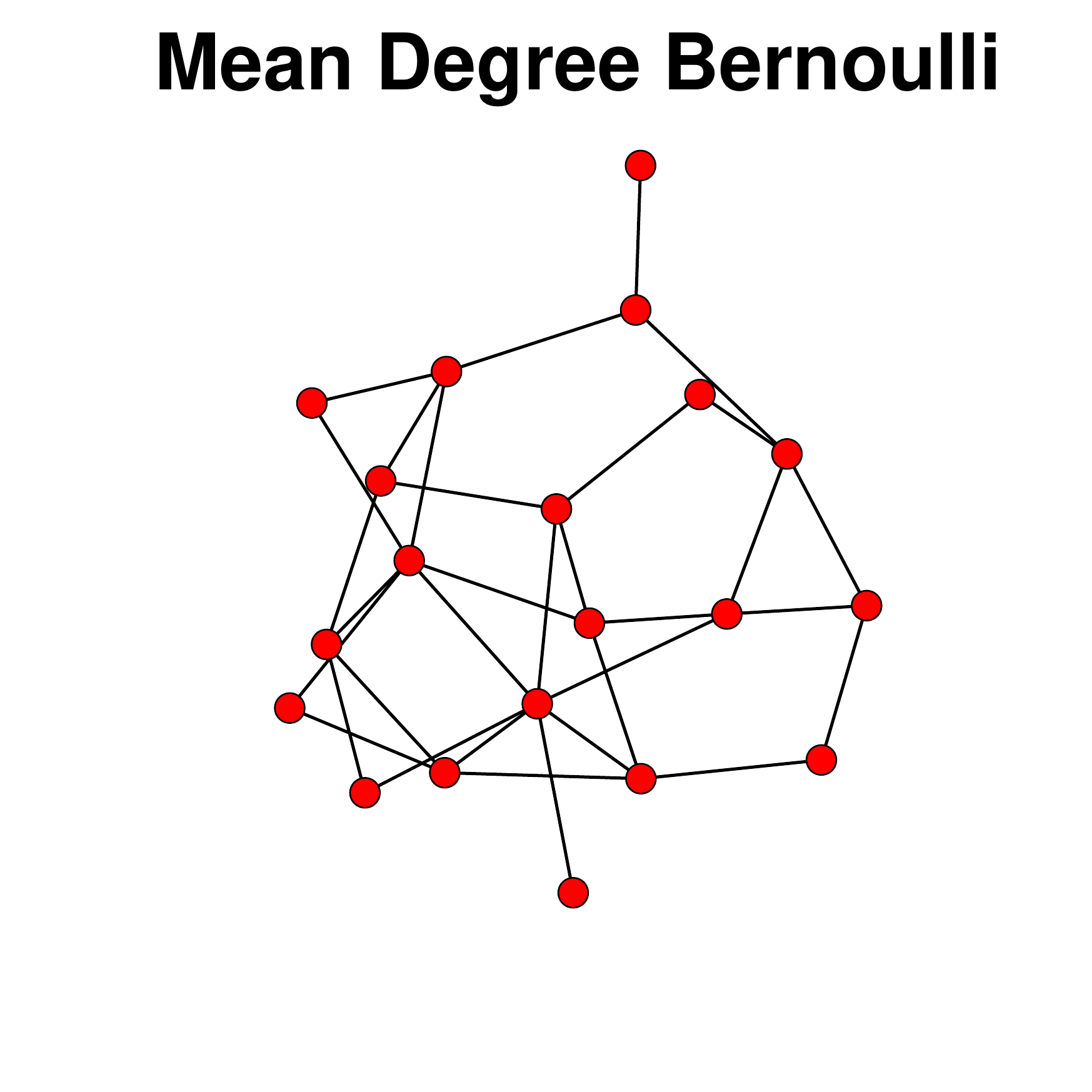}
\caption*{\scriptsize$e^{\theta}=3$}
\end{subfigure}
\begin{subfigure}{.15\textwidth}
\includegraphics[width=.95\linewidth]{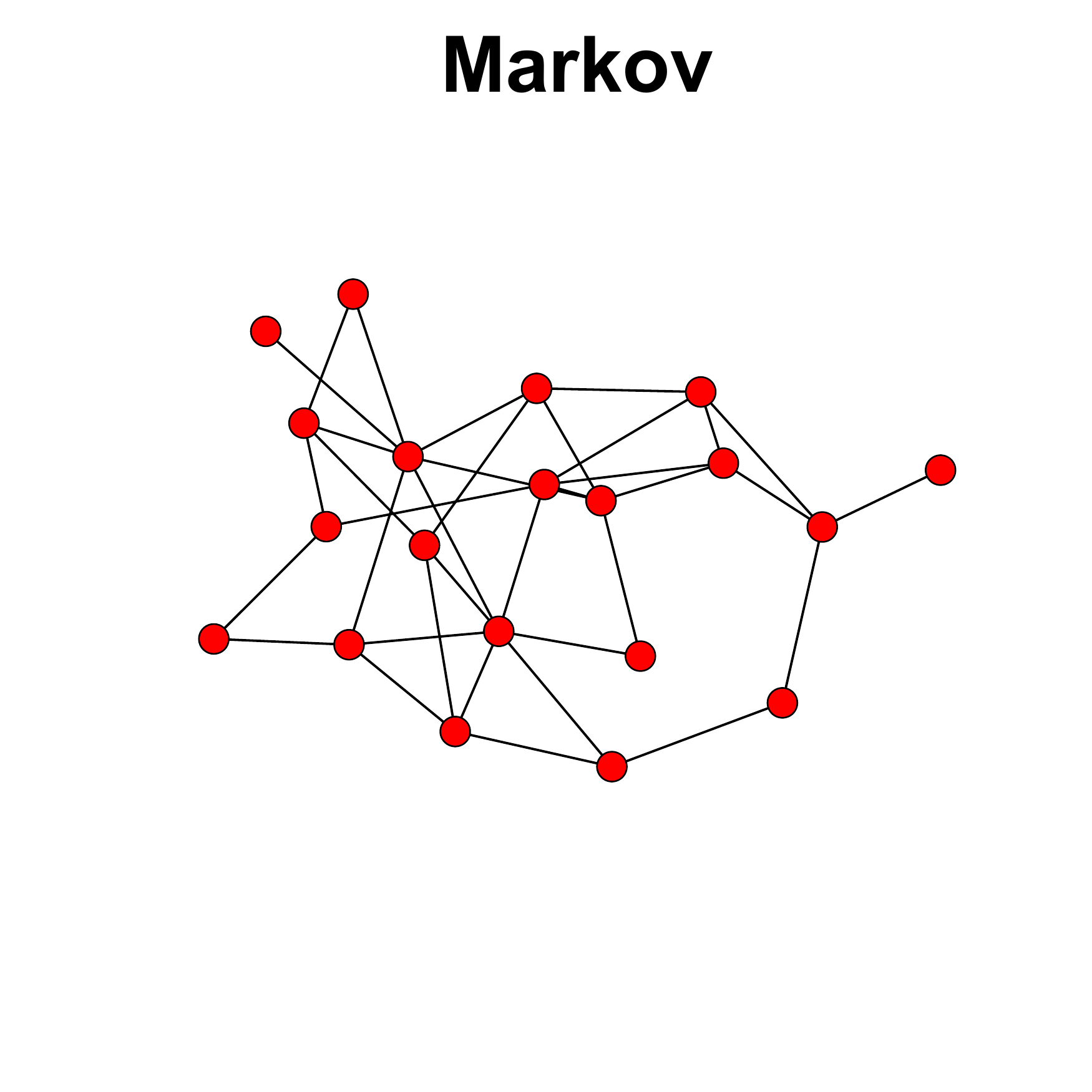}
\caption*{\scriptsize$\bm{\theta}= \{ -1.55, -0.05, 0.25 \}$}
\end{subfigure}
\begin{subfigure}{.15\textwidth}
\includegraphics[width=.95\linewidth]{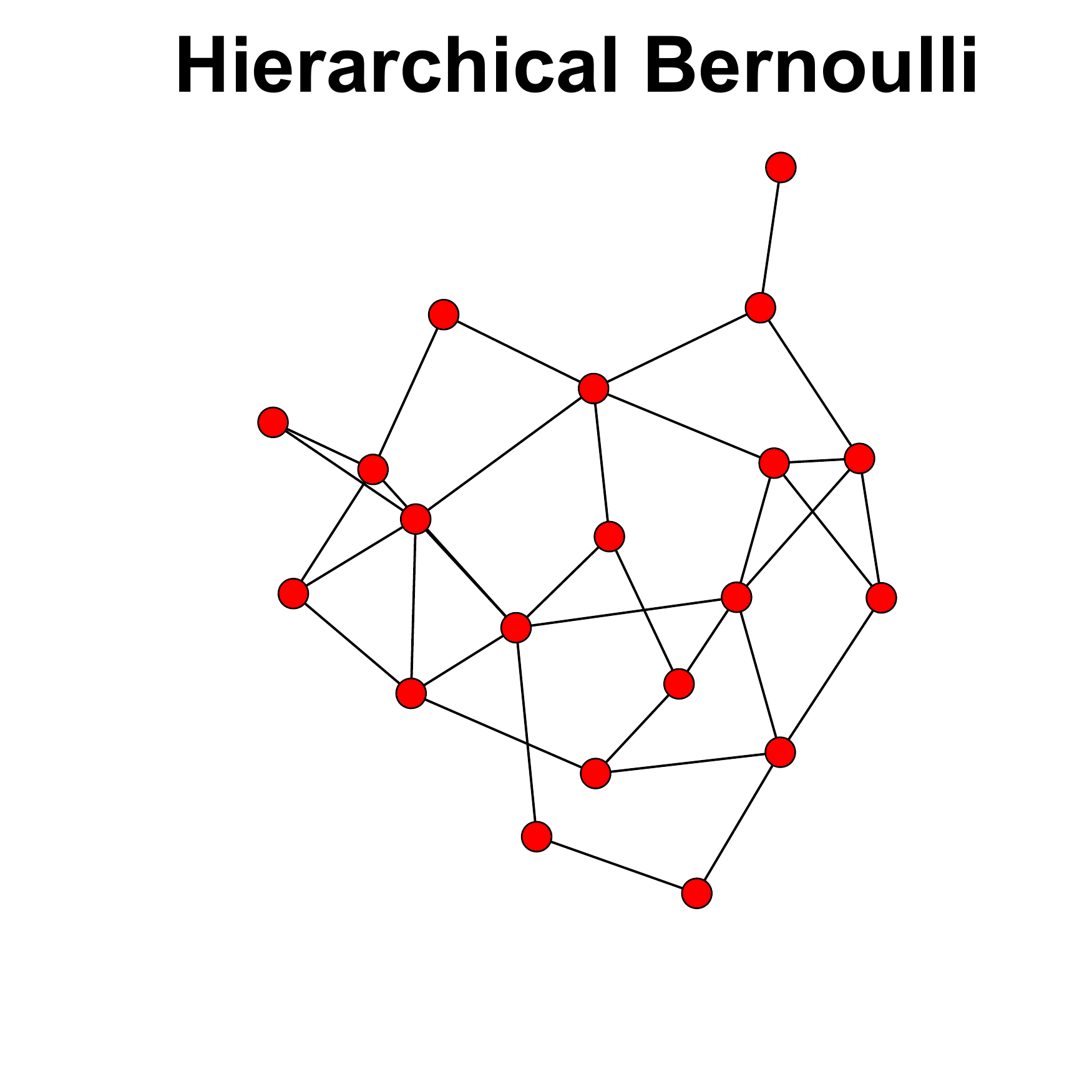}
\caption*{\scriptsize$p_{within}=0.20$, \\ $p_{btw}=0.10$}
\end{subfigure}
\begin{subfigure}{.15\textwidth}
\includegraphics[width=.95\linewidth]{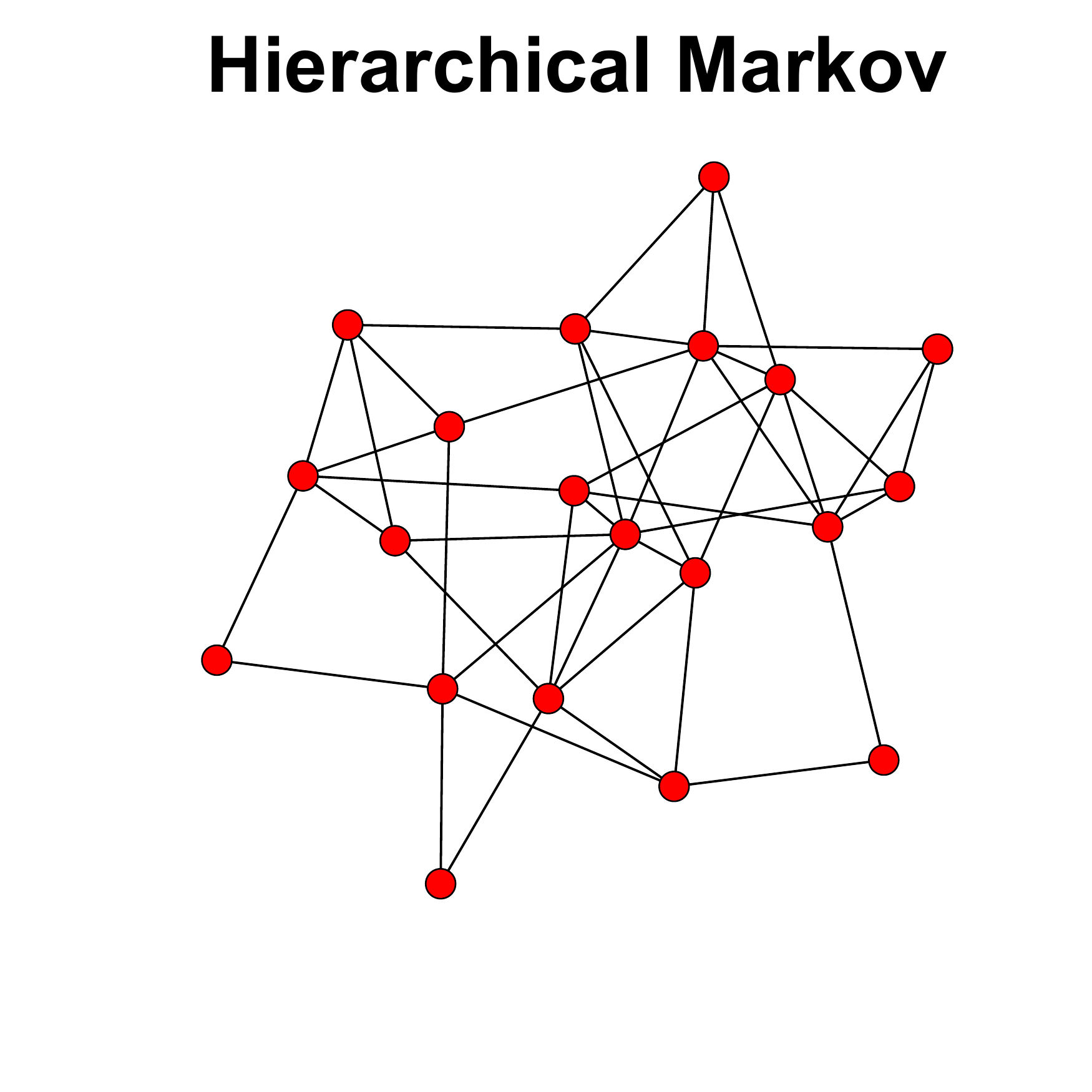}
\caption*{\scriptsize$\bm{\theta}=\{ -1.55, -0.05, 0.25 \},$ \\ $p_{btw}=0.25$}
\end{subfigure}
\label{fig:simDat}
\end{figure*}

\sloppypar{\textbf{Markov ERGM}.  Exponential random graph models (ERGMs) provide an easy way to introduce dependence between tie variables.  Here, we use the Markov version of an ERGM which specifies that all tie variables which share a node are dependent \citep{frank_strauss_1986}.  The triad version of this model is formulated as follows:
\begin{align}
P( \bm{Y} = \bm{y} | \bm{\theta} ) &= \text{exp} \left\{ <\bm{\theta},\bm{h}(\bm{y}) > - \psi(\bm{\theta}) \right\} 
\end{align}
where $<\bm{x}_1,\bm{x}_2>$ denotes the inner product of $\bm{x}_1$ and $\bm{x}_2$, the elements of $\bm{h}(\bm{y})$ are counts for the number of edges, 2-stars, and triangles in the observed network $\bm{y}$, respectively, i.e. $\bm{h}(\bm{y}) = \left\{ h_{edge}(\bm{y}), \> h_{2-star}(\bm{y}), \> h_{triangle}(\bm{y}) \right\}$, $\bm{\theta}$ are the corresponding parameters, i.e. $\bm{\theta} = \left\{ \theta_{edge}, \> \theta_{2-star}, \> \theta_{triangle} \right\}$, and $\psi(\bm{\theta})$ is the normalizing constant:}
\[ \psi(\bm{\theta}) = \sum_{\bm{y}^* \in \mathcal{Y}} \text{exp} \left\{ <\bm{\theta},h(\bm{y}^*) > \right\}. \]

We base parameter values for our generative Markov ERGM off of the well-known Florentine marriage network \citep{padgett_1994}\footnote{This network was chosen arbitrarily, except for the requirement that the triad Markov ERGM could be successfully fit to it.  Our primary goal is to create simulated networks that in some way resemble real world networks, but not to recreate the dynamics of this particular network per se.}.  Specifically, we fit the triad model to this network and use the resulting parameter estimates for our simulations:  $\theta_{edge} = -1.55, \> \theta_{two-star} = -0.05, \>\text{and } \theta_{triangle} = 0.25$.  This allows us to represent some of the structural and topological features that we often observe in real world networks.  However, there are many well-known issues with ERG models, particularly the fact that all nontrivial specifications form non-projective families \citep{shalizi_rinaldo_2013}.  This indicates that parameter values do not have the same meaning across networks of different size.  Yet we include the Markov ERGM here due to its widespread popularity in network analysis.

Note that we do not incorporate any of the latest modifications to ERGM statistics, such as the geometrically weighted terms developed by \citet{snijders_pattison_robins_2006}.  These are largely solutions to model-fitting issues and not necessarily required for the generative models we specify here.  Further, the relationship between these tuning parameter values and network size is unclear and potentially complicated.

\textbf{Hierarchical Bernoulli Model}.  Both the Hierarchical Bernoulli Model and hierarchical Markov ERGM are special cases of the more general HERGM, or Hierarchical ERGM \citep{schweinberger_handcock_2015}.  Generally, this model utilizes a Bayesian approach to identify plausible partitions of the network into smaller sub-networks (or ``neighborhoods'') and enforces strong dependence (e.g. a Markov ERGM) within neighborhoods with weak dependence (i.e. the Bernoulli model) between neighborhoods.  In this way, this model avoids the ERGMs' issue with non-projectability.  Thus, we consider these generative models our best attempt to mimic real world networks that vary in size.

For both of our HERGM simulated datasets, we consider the number of neighborhoods, $K$, to be fixed (although neighborhood membership, $\bm{Z}$, is still assumed unknown) and we set $K = \frac{n}{5}$ so that neighborhood size remains relatively small while the number of neighborhoods varies smoothly with network size.  Let $\bm{Y}_{k,k}$ denote the subgraph of $\bm{Y}$ corresponding to the $k$th neighborhood and let $\mathcal{A}_k$ represent the set of nodes belonging to the $k$th neighborhood.  Then, the probability distribution for $\bm{Y}$ can be written as follows:  
\begin{align}
P( \bm{Y} = \bm{y} | \bm{Z}, \bm{\theta} ) &= \prod_{k=1}^K P(\bm{Y}_{k,k} = \bm{y}_{k,k} | \bm{Z}, \bm{\theta}) \times \prod_{l=1}^{k-1} \prod_{i \in \mathcal{A}_k, j \in \mathcal{A}_l} P( Y_{ij} = y_{ij} | \bm{Z}, \bm{\theta} ) \nonumber
\end{align}
where the node membership vectors are dsitribtued as $\bm{Z}_i | \bm{\pi} \overset{iid}{\sim} \text{Multinomial}( 1; \bm{\pi} )$ for $i=1,...,n$.

Notice that the probability distribution for $\bm{Y}$ is conditional on the model parameters as well as the membership vectors: the set of $\bm{Z}_i$'s.  From (3), we can see that this model induces local dependence, since the probability mass function factorizes into within-neighborhood and between-neighborhood pieces.  Further, as suggested by \citeauthor{schweinberger_handcock_2015}, we make the following simplifying assumptions:  First, we assume that the between-neighborhood probability mass function depends only on the presence of edges:
\[ P(Y_{ij}=y_{ij} | \bm{Z}, \bm{\theta})  = \text{exp} \left\{ \theta_{btw} y_{ij} - \psi(\theta_{btw}) \right\} \]
and secondly, that the within-neighborhood probability mass function resembles a generic ERGM:
\begin{align}
P( \bm{Y}_{k,k} = \bm{y}_{k,k} | \bm{Z}, \bm{\theta} ) = \text{exp} \left\{ <\bm{\theta}_W, \> \bm{h}_W (\bm{y}_{k,k})> - \psi(\bm{\theta}_W) \right\}
\end{align}
where $\psi(\theta_{btw})$ and $\psi(\bm{\theta}_W)$ are normalizing constants, calculated similarly as for Equations (1) and (2).

As suggested by \citeauthor{schweinberger_handcock_2015}, we specify a multivariate normal prior for $\bm{\theta}$ with parameters $\bm{\mu}$ and $\bm{\Sigma}$, and chose a Dirichlet prior for $\bm{\pi}$, with parameter $\alpha \bm{1}_K$, where $\bm{1}_K$ is a $1\times K$ vector of ones.  For our simulations, we set $\alpha=10$ and consider $\theta_{btw}$ fixed.  For all models we consider elements of $\bm{\theta}$ uncorrelated and without loss of generality set the diagonal elements of $\bm{\Sigma}$ equal to one.   

For the Hierarchical Bernoulli Model, the within-neighborhood ERGM depends only on tie density (i.e. $\bm{h}_W (\bm{y}_{k,k})$ in Equation (4) consists of an edge count for the $k$th neighborhood subgraph of $\bm{y}$).  So, for this model, note that $\bm{\theta}_W$ is one-dimensional and we use the value corresponding to a within-neighborhood density of 0.20.  We set $\theta_{btw}$ so that the probability of connections between neighborhoods is 0.10.

\textbf{Hierarchical Markov ERGM}.  This is another special case of the HERGM, where $\bm{h}_W (\bm{y}_{k,k})$ is composed of subgraph counts for edges, 2-stars, and triangles, very similar to the setup for the Markov ERGM (see Equation (2)).  We use the same parameter values as in our traditional Markov ERGM - those estimated from the Florentine marriage network.  That is, we set $\bm{\theta}_W = \{ \theta_{edge}, \>\theta_{2-star}, \>\theta_{triangle} \} = \{ -1.55, \>-0.05, \>0.25 \}$ to govern the Markov ERGM within neighborhoods.  We set $\theta_{btw}$ so that the probability of connections between neighborhoods is 0.25.

Examples of realizations from these generative network models are shown in Figure \ref{fig:simDat}.  We plot randomly selected networks of the smallest size examined here, $n=20$, so that structural differences are more readily visible.


\subsection{Direct Comparison} \label{sec:dircomp}

Recall that for each of these generative models, we simulate $N_n=200$ replicates with $n = 20, 30, ... ,100$ nodes.  All networks are simulated using standard functions from the \texttt{statnet} \citep{handcock_hunter_butts_etal_2015} and \texttt{hergm} \citep{schweinberger_handcock_luna_2015} packages for \texttt{R}.  For each network in each of these simulated datasets, we compute each of the common network statistics described in Section \ref{sec:stats}.  Then we consider the distribution of each of these common network statistics where the networks come from the same generative model, paying specific attention to how the distribution varies according to the size of the network.  To examine this, consider the following three metrics.

\begin{figure*}[t]
\caption{Histograms of Unadjusted Statistics}
\centering
\includegraphics[width=.9\textwidth]{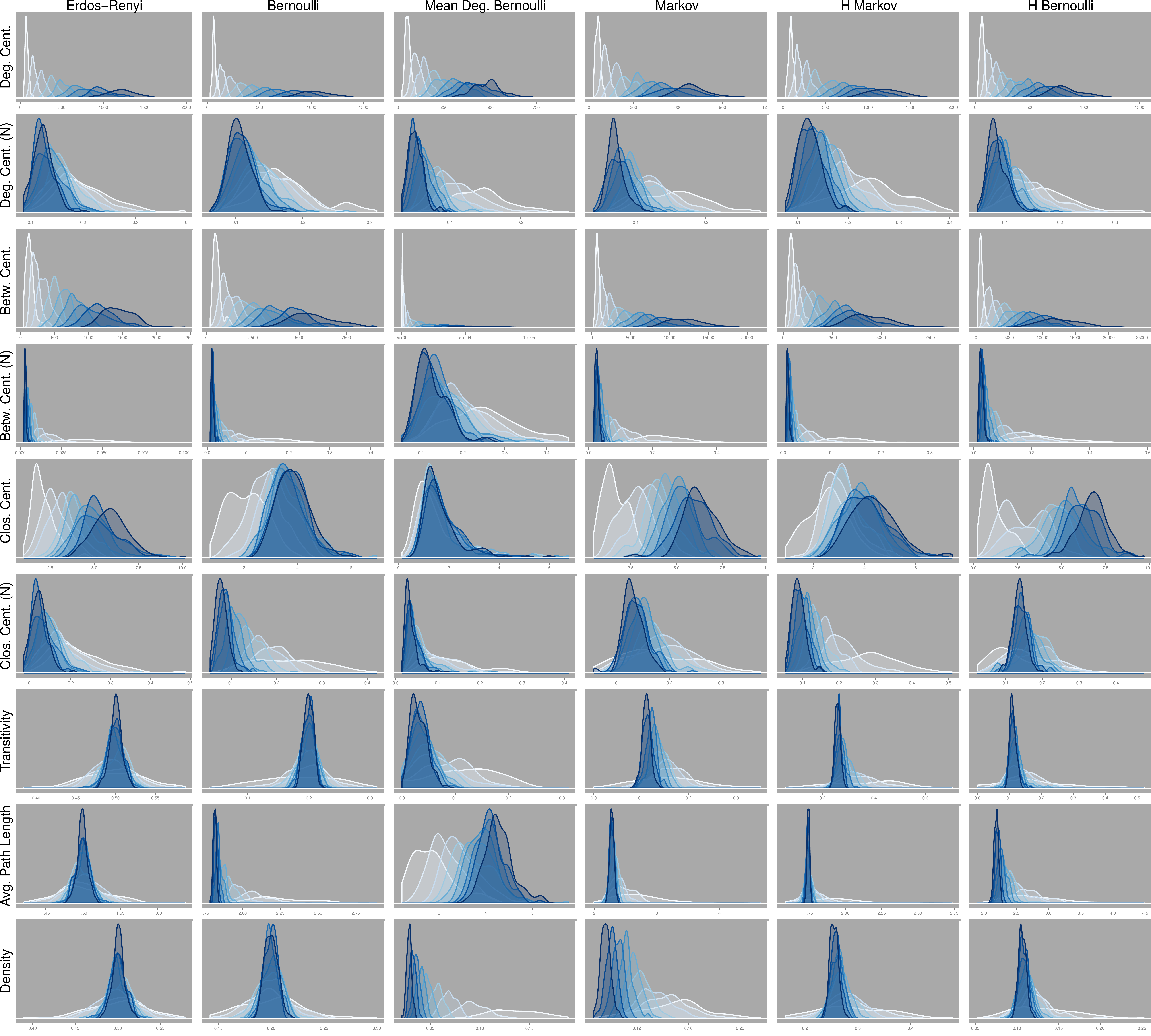}
\\
\includegraphics[width=.35\textwidth]{Figure4_legend.png}
\vspace{1mm}
\caption*{The columns indicate different simulated datasets while the rows represent the different network statistics of interest.}
\label{fig:Hist_Unadj}
\end{figure*}

First, perhaps the most intuitive method for examining differences in distributions is using histograms.  In Figure \ref{fig:Hist_Unadj}, we have plotted the histograms for each common network statistic under each generative model.  In each plot, the color of the histogram indicates the size of the network, with the smallest network size ($n=20$) shown in white and the largest network size ($n=100$) shown in dark blue.  Ideally, we would like the histograms to stack up neatly on top of each other with little difference in overall shape.

\begin{figure*}[t]
\caption{Heat Maps of Kolmogorov-Smirnov Statistics for Unadjusted Statistics}
\centering
\begin{minipage}[position=t,inner-pos=t]{.45\textwidth}
  \includegraphics[width=.95\linewidth]{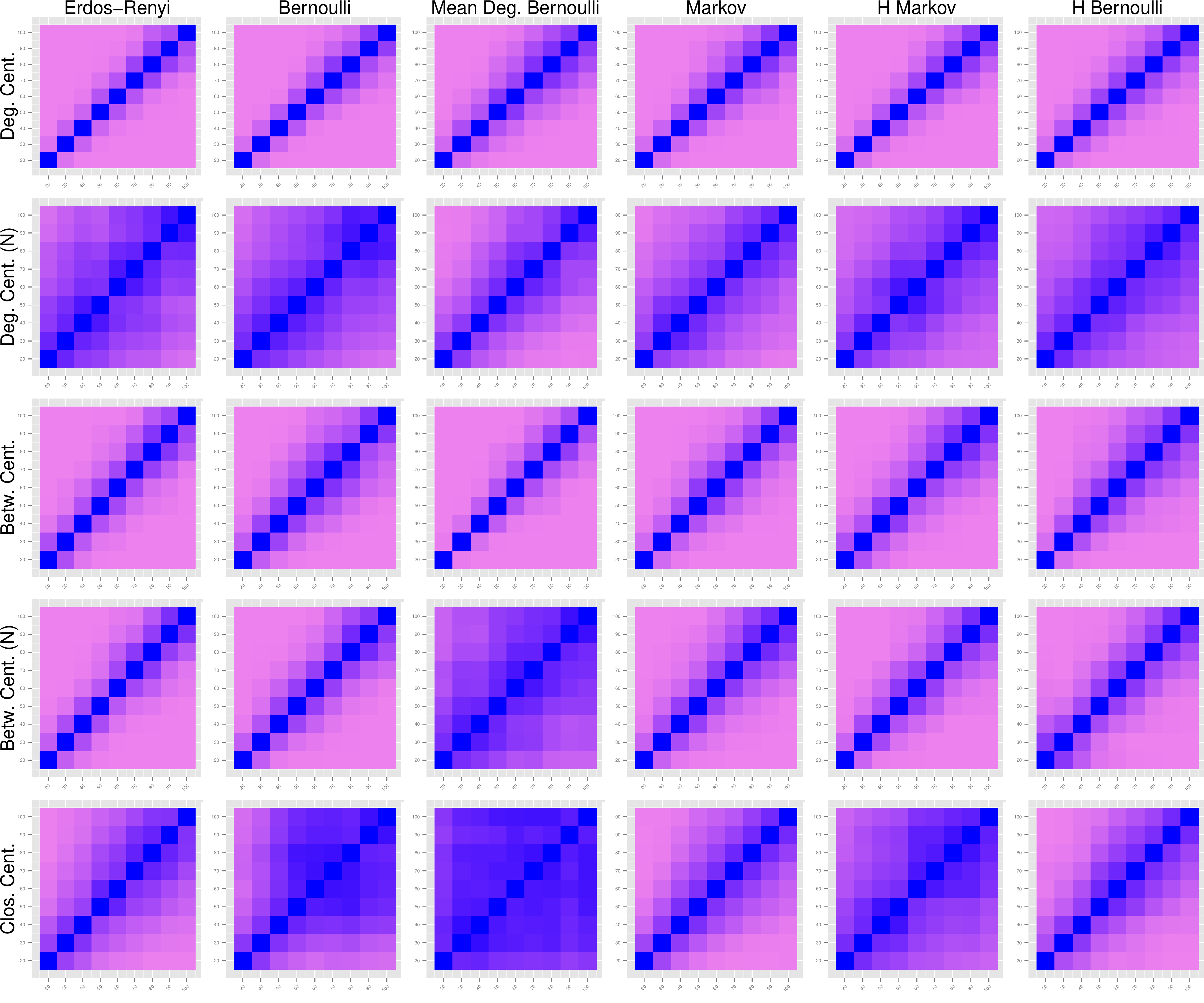}
\end{minipage}
\hspace{-2mm}
\vrule
\hspace{2mm}
\begin{minipage}[inner-pos=t]{.45\textwidth}
  \includegraphics[width=.95\linewidth]{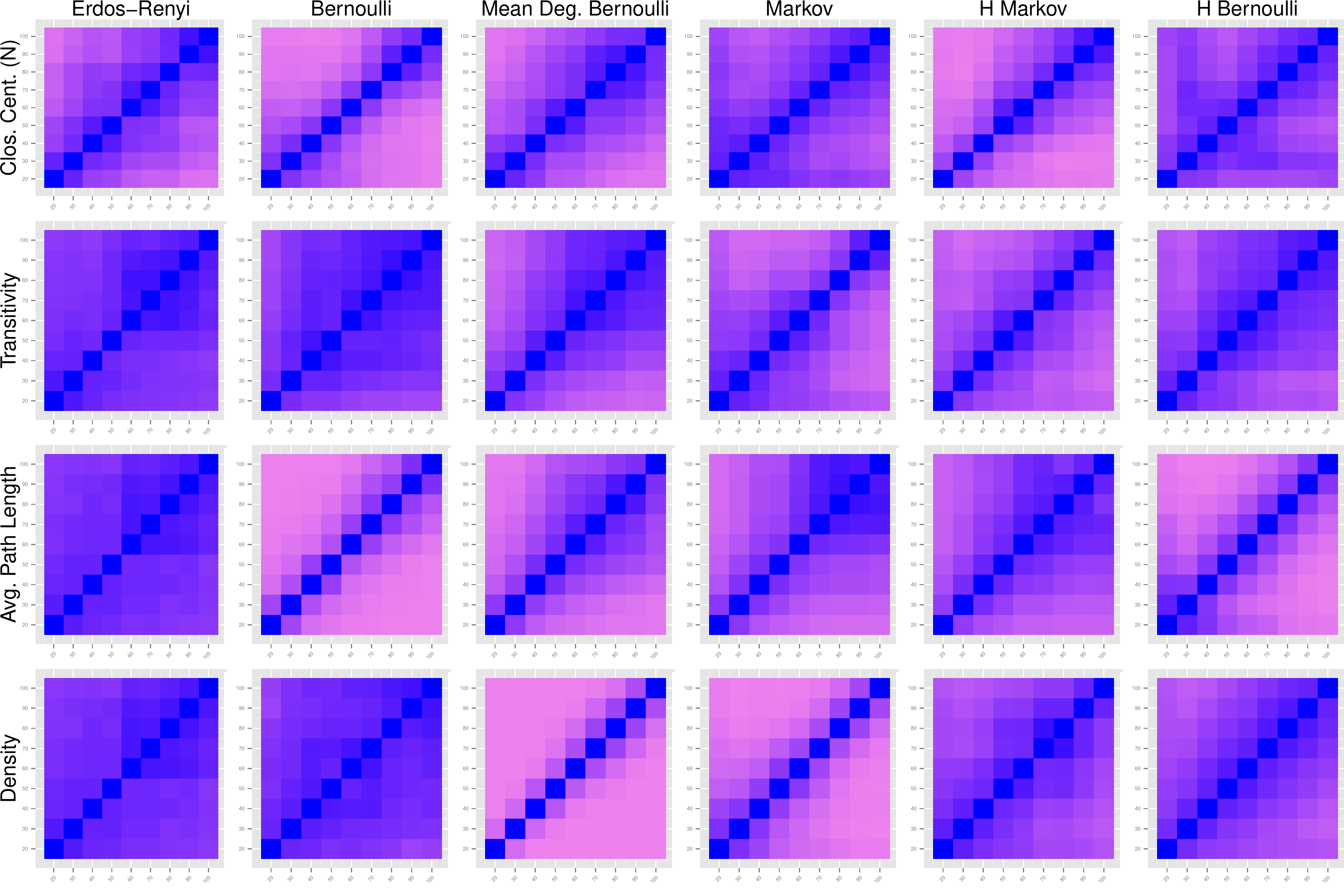}
  \vspace{.5in}
\end{minipage}
\begin{minipage}[inner-pos=c]{.05\textwidth}
  \centering
  \includegraphics[width=.75\linewidth]{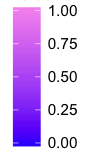}
\end{minipage}
\caption*{\\The columns indicate different simulated datasets while the rows represent the different network statistics of interest.  Each cell in a plot represents the 2-sample Kolmogorov-Smirnov statistic for the comparison of a network statistic's distribution across two different network sizes.  The network sizes are indicated on both the x- and y-axes.  Cells that are dark blue indicate that the distributions are very similar while those that are violet indicate that the distributions differ.}
\label{fig:KSstat_Unadj}
\end{figure*}

Second, we attempt to quantify the difference in these distributions by considering 2-sample Kolmogorov-Smirnov statistics.  Generically, let $F_{M,\eta,n_i}$ denote the empirical distribution function (EDF) for a network statistic, $\eta(G)$, calculated on networks from Model $M$ of size $n_i$, for $M=1,...,5$ denoting the generative models discussed in Section \ref{sec:models}, $\eta(G)$ from the set of common network statistics described in Section \ref{sec:stats}, and $i=1,...,9$ corresponding to $n_i=20,30,...,100$.  Note that for all $M, \eta,$ and $n_i$, the number of samples from $F_{M,\eta,n_i}$ will be the same, precisely $N_n=200$.  We use the Kolmogorov-Smirnov statistic to compare empirical distributions of a statistic calculated on networks from the same model across two different network sizes:
\[ \sup_{x \in \mathbb{R}} | F_{M,\eta,n_i}(x) - F_{M,\eta,n_j}(x) | \hspace{15mm} i \ne j.\]
This method allows us to compare only two network sizes simultaneously, so we calculate this statistic pairwise for each possible pair of network sizes.  The Kolmogorov-Smirnov statistic is naturally bounded between zero and one, and so we can easily compare across different models and network statistics.  In Figure \ref{fig:KSstat_Unadj}, we present heatmaps where the cells of each heatmap correspond to the Kolmogorov-Smirnov statistic computed as the difference between empirical distributions for networks of two different sizes.  Dark blue cells correspond to distributions that are very similar, while violet cells correspond to distributions that are quite different.  Note that these plots are symmetric, so it is sufficient to examine only the upper left or lower right triangle.

Third, we consider further reducing our summary information by computing k-Sample Anderson-Darling statistics \citep{scholz_stephens_1987}.  For a particular generative model and network statistic, we use this k-Sample statistic to summarize the difference across distributions of all network sizes simultaneously.  The k-Sample Anderson-Darling statistic is formulated as follows:
\[ \sum_{i=1}^9 N_n \int_{B_{M,\eta}} \frac{ \{ F_{M,\eta,n_i}(x) - H_{M,\eta}(x) \}^2 }{H_{M,\eta}(x) \{ 1 - H_{M,\eta}(x) \} } \]
where $H_{M,\eta}(x) = \sum_{i=1}^9 F_{M,\eta,n_i}(x) $ is the empirical distribution function of the pooled sample across all network sizes and $B_{M,\eta} = \{ x \in \mathbb{R} : H_{M,\eta}(x) < 1\}$.  Note that the typical equations simplify here since, as mentioned previously, the number of samples from $F_{M,\eta,n_i}$ is the same for all $M, \eta,$ and $n_i$.  The Anderson-Darling statistic is a quadratic empirical distribution function statistic (the Cramer-von Mises statistic is another special case) and measures the weighted distance between the empirical distributions under consideration.  The k-Sample Anderson-Darling statistics for our network data are provided in Table \ref{tab:kSamp_Unadj}.

\begin{table*}[t] 
\caption{k-Sample Anderson-Darling Statistics for Unadjusted Statistics}
\footnotesize
\centering
\begin{tabular}{c|l||rrrrrr}
\multicolumn{1}{l}{} && \multicolumn{6}{c}{\textsc{Simulated Dataset}}\\
  \cline{3-8}
\multicolumn{1}{l}{} && \multirow{2}{*}{\ER} & \multirow{2}{*}{Bernoulli} & Mean Degree & \multirow{2}{*}{Markov} & Hierarchical & Hierarchical \\
\multicolumn{1}{l}{} && & & Bernoulli & & Markov & Bernoulli \\ 
  \hline
  \hline
\parbox[t]{2mm}{\multirow{9}{*}{\rotatebox[origin=c]{90}{\textsc{Statistic}}}} & Deg. Cent. & 694.16 & 680.85 & 609.90 & 642.89 & 667.88 & 638.90 \\ 
  & Deg. Cent. (N) & 209.65 & 194.63 & 391.44 & 329.30 & 266.07 & 226.86 \\ 
  & Betw. Cent. & 633.16 & 580.17 & 694.14 & 639.21 & 591.44 & 600.95 \\ 
  & Betw. Cent. (N) & 637.46 & 609.14 & 179.13 & 568.19 & 617.07 & 564.86 \\ 
  & Clos. Cent. & 409.44 & 215.38 & 31.77 & 465.80 & 192.43 & 467.63 \\ 
  & Clos. Cent. (N) & 229.20 & 452.93 & 304.60 & 146.13 & 439.05 & 130.31 \\ 
 & Transitivity & 67.84 & 81.18 & 186.36 & 238.35 & 225.82 & 121.85 \\ 
  & Avg. Path Length & 60.09 & 558.74 & 356.51 & 228.62 & 171.51 & 430.91 \\ 
  & Density & 58.08 & 57.62 & 682.59 & 533.56 & 127.57 & 132.75 \\ 
\end{tabular}
\label{tab:kSamp_Unadj}
\end{table*}

We see that in most cases, statistics across networks of different sizes clearly have very different distributions and comparing them directly is generally not appropriate.  Not surprisingly, in many of the heat maps and histogram plots, we notice that direct comparison is least appropriate when the difference between network sizes is large.  Also notice that, given a fixed difference in network sizes, direct comparison appears to be less appropriate when the network sizes themselves are small.  For example, comparing a network statistic across networks of sizes $n=20$ and $n=30$ appears to be more problematic than across networks of sizes $n=90$ and $n=100$.  This is not surprising, since in the former case the difference in network size constitutes a much larger proportion of the absolute network sizes themselves.

However, there are cases where directly comparing network statistics is not terribly inappropriate.  For example, the distributions of the topological measures appear to be rather comparable across different sizes.  Yet, this statement does not appear to hold (nearly as well) if we are considering average path length in a group of Bernoulli graphs or density from a group of Markov graphs.  Of course, in practice, the generative model is unknown.  Thus, we would advise avoiding direct comparison if possible, even for these relatively stable topological measures.

In terms of the centralization measures, we notice that using the normalized versions appears to greatly lessen the issue, though it is not entirely solved.  Notably, in the case of betweenness, normalization appears to be less effective (this is particularly evident in the heat maps in Figure \ref{fig:KSstat_Unadj} and the Anderson-Darling statistics in Table \ref{tab:kSamp_Unadj}), except for networks from the mean degree preserving Bernoulli model.  Further, in the case of closeness centrality, normlization appears to offer improvements \textit{only} in combination with particular types of graphs.  Particularly, if we examine the 2-Sample Kolmogorov-Smirnov statistics in Figure \ref{fig:KSstat_Unadj} for the two versions of the closeness centrality measure, we see that the normalized version is superior only in the cases where the simulated data are generated from an \ER model, Markov ERGM, or Hierarchical Bernoulli model.  In the other simulated datasets, the unnormalized statistics are more comparable across network sizes.  Perhaps the normalization for closeness centrality is sensitive to particular types of topological network structure.  However, it is unclear from these preliminary results what the sensitivity might be and/or how to adjust for it.

In short, this simulation study has clearly demonstrated that across a range of popular and reasonable generative network models, network statistics are not directly comparable across networks of different size, even when the networks under consideration are generated from the same model.  It follows then that statistics calculated on real world networks (say from the same data generating process) are not directly comparable either.  Thus, we turn to a brief literature review of proposed adjustments to our common network statistics which should allow for comparison across networks of different size.


\section{Comparing Network Statistics}

\subsection{Review of Existing Approaches} \label{sec:rev}

\citet{anderson_butts_carley_1999} suggest specifying a baseline network model and using the distribution of the desired statistic calculated on a set of networks simulated from this model as a reference distribution, which can be used to standardize the observed statistic.  Based on the observation that many common network statistics depend on both the size and density of the underlying network, they suggest using a model which adjusts for these characteristics.  However, when comparing networks that vary in both size and density, this means making comparisons across networks based on reference distributions that also vary across the networks.  Instead, comparisons would be more straight forward and more readily interpretable if the comparisons across networks are made relative to a single reference distribution that is common across all the networks under consideration and represents some kind of absolute baseline model.  In this sense, not only can one compare networks within a dataset but across datasets where the adjustments were perhaps performed by different researchers.  Of course, choosing a realistic fully parameterized reference distribution for comparison a priori is certainly not a straight forward task.  A simple choice might be a completely random graph, or the \ER model.  We investigate this option as a special case of our more general method in Section \ref{sec:Sim}.  Intuitively, standardizing relative to the \ER model might be an improvement over direct comparison, but could remain problematic in some cases since \ER networks simply fail to exhibit the type of structural detail and topological features that we typically observe in real world networks.  Of course, we could consider different versions of an a priori fully parameterized reference distribution, but as \citet{vanwijk_stam_daffertshofer_2010} point out, this requires complete trust in the validity of whichever baseline model is used.


\subsection{A New Mixture Model Reference Distribution}

We propose a new reference distribution which is formed by a mixture of random graph models, where we mix over the dependence and structural features actually manifested in the observed dataset.  Specifically, we suggest mixing over the dependence structure exhibited in a subset of networks randomly selected from the entire collection of observed networks.  In this way, the reference distribution is common across all networks being considered and will, by design, mimic the structure of the networks being compared as well as reflect the natural variability among the observed networks.  Our proposed methodology hopes to draw on the historical success of mixture models as a class of flexible models which are particularly advantageous in situtations where it is difficult or undesirable to fully specify a traditional model a priori, such as in \citet{newcomb_1886}'s model for outliers or \citet{pearson_1894}'s work with evolutionary populations.  However, note that our proposed methodology will result in \textit{relative} comparisons across networks, since the reference distribution is empirically-based rather than formed a priori.

This type of relative comparison is not unique to the methodology we propose here.  Consider the familiar case in which a set of correlated variables are considered for inclusion as covariates in a regression analysis.  Including the full set directly would lead to unstable coefficient estimates due to multicollinearity.  Instead, a dimension-reduction technique like principal components analysis (PCA) can be used; PCA uses an orthogonal transformation to convert a set of possibly correlated variables into a set of linearly uncorrelated transformations of these variables, called principal components.  And it does so in such a way that the first principal component accounts for the largest variance in the original set of variables, the second principal component accounts for the second largest amount of variance and so on.  Although the value of the principal components themselves do not have the same physical meaning as the orginal set of variables which were first considered, they still capture (most of) the variability and the underlying content of the original set.  Similarly, the value of a network statistic adjusted via the mixture model adjustment will not have meaning in and of itself, but will be very meaningful relative to statistics calculated on other networks that are adjusted in the same way.  Thus, similar to a PCA style analysis, results from the mixture model adjustment will pick out differences among the set of observed networks, rather than relative to some absolute scale. 

The procedure for this method can be divided into two main components:  Again, suppose we observe a collection of networks $ \bm{Y} = \left\{ \bm{Y}_1, \bm{Y}_2, ... \bm{Y}_N \right\}$ with corresponding sizes $ \bm{n} = \{n_1, n_2, ... n_N \}$.  Let $\eta( \cdot )$ be our network statistic of interest.

\vskip1.5ex
\noindent \textit{Component 1:  Mixture Model Parameter Selection}\\[-.3in]
\begin{enumerate}
\item Randomly select $N_M$ networks from$\> \bm{Y}$
\item Fit a graph model separately to each of these networks.  Let the obtained parameter estimates be denoted by
\[ \widehat{\bm{\theta}} = \left\{ \widehat{\bm{\theta}}_1, \widehat{\bm{\theta}}_2, ... \widehat{\bm{\theta}}_{N_M} \right\} \]
where $\widehat{\bm{\theta}}_j$ is the vector of parameter estimates from fitting the model to the $j$th network selected in step one.
\end{enumerate}

\vspace{1mm}
\noindent \textit{Component 2:  Mixture Model Simulation and Adjustment }\\[.1in]
For $i=1,...N$,
\begin{enumerate}
\item For $j = 1, ... N_M$, simulate $\frac{N_S}{N_M}$ graphs of size $n_i$ setting $\bm{\theta} = \widehat{\bm{\theta}}_j$.
\item Call the entire collection of simulated networks
\[\widetilde{\bm{Y}}^{(i)} = \left\{ \widetilde{\bm{Y}}^{(i)}_1,  \widetilde{\bm{Y}}^{(i)}_2, ...  \widetilde{\bm{Y}}^{(i)}_{N_S} \right\}. \]
\item Compute $\eta \left( \bm{Y}^{(i)}_k \right) $ for $k=1,...N_s$ and find the sample mean, $m_{\eta}^{(i)} $, and sample standard deviation, $s^{(i)}_{\eta}$, of this distribution.
\item Finally compute the z-score:
\[ z_i = z \left( \bm{Y}_i | \widetilde{\bm{Y}}^{(i)}, N_S, N_M \right)  = \frac{ \eta(\bm{Y}_i) - m_{\eta}^{(i)} }{ s^{(i)}_{\eta} }. \]
\end{enumerate}
where $N_M\le N$ and $N_S$ is some large positive integer divisible by $N_M$.  Note that the computational efficiency of this algorithm can be improved by letting $i$ index only the unique elements in $\bm{n}$.

Here, we have chosen to use a very intuitive standardization measure, a z-score, though other measures have been suggested \citep{vanwijk_stam_daffertshofer_2010}.  While distributions of network statistics for \ER  graphs are known in some special cases, the distribution of a given statistic for some arbitrary network model is generally unknown.  However, from past observations and Figure \ref{fig:Hist_Unadj}, a normal distribution might be a relatively well-fitting approximation in some cases, and so, a z-score is a plausible standardization measure here.  

Notice that in this algorithm, we have intentionally left the form of the network model used for simulating the reference distribution in this procedure unspecified.  There may be situtations in which a researcher thinks one model is appropriate given the type of networks under consideration, or perhaps certain network statistics are more amenable to adjustments utilizing a particular class of network models.  One particular special case is worth pointing out:  If the \ER model is utilized, note that Component 1 of the above algorithm is no longer necessary since $\hat{\bm{\theta}}_j = \text{logit}(0.50) \>\>\forall j=1,...N_M$ and the reference distribution is not only common across all networks but also an absolute reference distribution (i.e. it is no longer a mixture distribution and does not depend on the data in any way).  

Finally, note that this procedure requires fitting $N_M$ separate network models and thus network models in which obtaining parameter estimates can be more easily automated will be preferred.


\section{Simulation Study:  $n$-dependence} \label{sec:Sim}

\subsection{Methods}

To consider the efficacy of this method, we turn to a simulation study.  Treating the raw statistics from Section \ref{sec:dircomp} as our observed datasets, we apply this standardization method with $N_M=30$ and $N_S=1000$ to each network statistic and generative model pairing.  Note that for each generative model considered, $N=1800$ networks.  In terms of the network model used for simulating the mixture reference distribution in this adjustment, we consider four options:  the special case of a fully parameterized reference distribution utilizing the \ER model, the traditional Bernoulli model, the mean degree preserving Bernoulli model, and the Hierarchical Bernoulli model.  For each resulting version of the Mixture Model Adjustment, the network models will be fit using standard functions in the \texttt{statnet} \citep{handcock_hunter_butts_etal_2015} and \texttt{hergm} \citep{schweinberger_handcock_luna_2015} packages for \texttt{R}.

Recall that the \ER model is simply a special case of the Bernoulli model, where we fix the probability of a tie to be 0.50.  Thus, we might expect to see improvements over results from an \ER adjustment, if we simply allow the density of the graphs in the reference distribution to vary, in a way that matches the empirical distribution of densities of our observed networks.  This is precisely what we allow by utilizing the Bernoulli model in our Mixture Model Adjustment\footnote{Note that the term ``Mixture'' in the name for our proposed methodology refers to the distribution of simulated networks in step one of Component 2 of our algorithm.  These networks are simulated according to a mixture of different specifications of a particular network model where the components of this mixture are formed in step two of Component 1.  For example, we might simulate networks from a Bernoulli model, mixing over the log odds of a tie between -2, 1.4, and 0.2.  Note that our proposed methodology does not (currently) incorporate any of the so-called mixture models for random graphs \citep[e.g.][]{daudin_picard_robin_2008} in which a single network is modeled as a mixture of random graphs.}

Similarly, using the mean degree preserving version of the Bernoulli model will mix over graphs that mimic the average degree of our observed networks.  Since density and average degree are closely related, we might expect these two versions of the approach to perform similarly.  However, recall that the main difference between these two models is their behavior as network size increases:  the Bernoulli model maintains density, while adding \citet{krivitsky_handcock_morris_2011}'s offset preserves mean degree.  We might expect that whichever regime is present in the observed dataset, that it's corresponding model might prove a better fit for the mixture model adjustment procedure.

Perhaps the next more complicated graph model would be some ERG-type model which includes not only a density effect but also some structural statistics, such as triangles or k-stars.  However, due to \citeauthor{shalizi_rinaldo_2013}'s result, we do not believe that ERGM parameters have the same meaning across different network sizes (\citeyear{shalizi_rinaldo_2013}).  So, using an ERGM to simulate the mixture reference distribution would most likely be inappropriate.  Further, ERGMs are notoriously difficult to fit, particularly when model terms are exclusively structural, and thus further unsuited for this methodology.

As mentioned earlier, the Hierarchical ERGMs developed by \citeauthor{schweinberger_handcock_2015} do not face such issues.  In fact, the authors show that HERGMs do form a type of projective family, although this type of consistency is a weaker condition than that considered by \citeauthor{shalizi_rinaldo_2013} \citep{schweinberger_handcock_2015}.  However, HERGMs in their current implementation take much longer to fit than traditional ERGMs (hours rather than minutes, even for relatively small networks), though the resulting model fits are significantly more stable and reliable.  Since our proposed method requires fitting $N_M$ network models, we use only the Hierarchical Bernoulli version of the HERGM, rather than the more complex Hierarchical Markov model described in Section \ref{sec:models}.  In this implementation, when we fit the Hierarchical Bernoulli to the $N_M$ randomly selected observed networks, we assume the number of neighborhoods is unknown and use a nonparametric stick-breaking prior for the neighborhood membership parameters, $\bm{\pi}$.  We set the maximum number of neighborhoods to be $K =\frac{n}{5}$.  When we simulate the reference distribution formed from the parameter estimates obtained above, we follow the same model set up as for the simulated Hierarchical Bernoulli networks (described in Section 2.2), letting $\alpha=10$ and setting $K, \bm{\theta}, \theta_{btw},$ and the diagonal entries of $\bm{\Sigma}$ equal to the estimates from the model fitting step in the first component of the algorithm.


\subsection{Results}

\begin{figure*}[t]
\caption{Performance of Mixture Model Adjustments}
\centering
\includegraphics[width=.95\textwidth]{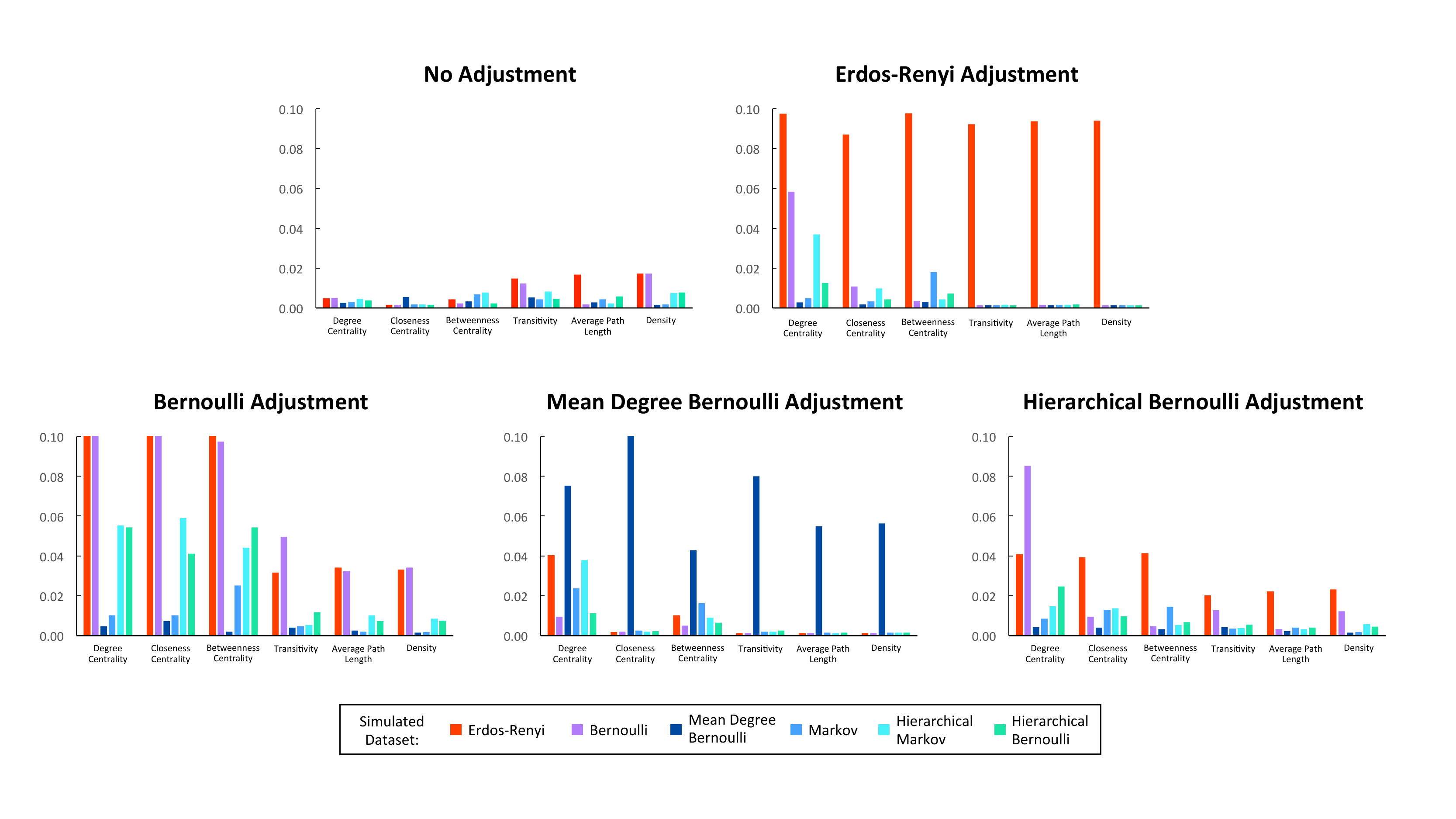}
\caption*{Each plot represents a different version of the mixture model approach, including no adjustment. The height of the bar corresponds to the inverse of the Anderson-Darling statistics, so that taller bars correspond to distributions that are more similar across varying network sizes.  For simplicityj, values above 0.10 are not shown here; the maximum is 0.13.}
\label{fig:BarChart}
\end{figure*}

As before, we use the three metrics from Section \ref{sec:dircomp} to examine the results of these mixture model adjustments:  histograms, heat maps of Kolmogorov-Smirnov statistics, and k-Sample Anderson-Darling statistics.  Intuitively, if the method is effective, then, relative to the comparison metrics for the raw statistics, we expect the histograms to overlap each other, the heat maps to have more dark blue cells, and the Anderson-Darling statistics to be smaller.  Note that the normalized versions of the centralization measures are excluded here since the normalization adjustment is linear for a fixed network size and thus z-scores for the normalized and unnormalized versions of these statistics are equivalent.  These figures are available in the Supplementary Materials.  As a summary of these results, we provide bar charts in Figure \ref{fig:BarChart} of the reciprocal of the Anderson-Darling statistics across each adjustment method, so that taller bars correspond to better performance (i.e. adjusted distributions of network statistics that are more similar across network sizes).

Not surprisingly, the \ER adjustment works strikingly well when the data are actually generated from the \ER model.  However, problems occur when the observed networks more strongly resemble real world network data, i.e. are simulated from the other generative models considered here.  In particular, this adjustment performs very poorly for all of the topological measures computed on non-\ER simulated networks.  However, we do see that using the \ER adjustment for the centralization measures appears to be an improvement over direct comparison.  Thus, the \ER adjustment method described here is certainly an improvement over direct comparison, but is still sometimes problematic (especially in the case of topological measures).  Note also that this adjustment method is very easily implemented, computationally inexpensive (for graphs similar in size to those examined here - the computational burden increases with network size), and is an example of a fully parameterized reference distribution.

Considering the first nontrivial version of the Mixture Model Adjustment, where the Bernoulli model is used, we see many improvements relative to adjustment using the \ER model.  Recall that the \ER adjustment seemed most suited for centralization scores.  However, we see even further improvements in the results based on this Bernoulli version of the Mixture Model Adjustment.  The results from this mixture model adjustment are more comparable across \textit{all} simulated datasets.

Recall also that the \ER adjustment performed quite poorly for the topological measures, so it is more appropriate to compare the results of the Mixture Model Adjustments to the raw topological network statistics themselves.  Here we see significant improvements, particularly for the simulated datasets where the networks are generated from the Bernoulli model and both versions of the HERGMs implemented here.

However, the Bernoulli version of the Mixture Model Adjustment for the network statistics calculated on Markov graphs does not appear to offer any advantanges over direct comparison of the statistics themselves.  This is not too alarming, given the previously mentioned result which shows that ERG-type models like this one do not form projective families \citep{shalizi_rinaldo_2013}.  This result implies that parameters in such models are not comparable across different network sizes and so, the dataset of simulated Markov graphs that we have used here might not be particularly meaningful, i.e. it might not exhibit the same type of structure across different network sizes.  If this is the case, we would not expect any type of adjustment method to be able to ``fix'' such an issue.

The simulation results examined here demonstrate that our proposed mixture model adjustment, where we use the Bernoulli model to form the simulated mixture reference distribution, offers a computationally inexpensive way to make reliable comparisons of both centralization and topological network statistics across networks of different size (see Table 5 in the Appendix for a comparison of computation times across methods).

As expected, the version of the Mixture Model Adjustment which uses the mean degree preserving Bernoulli model performs well for datasets where network degree is preserved as network size grows, i.e. on networks generated from the mean degree preservng Bernoulli model.  However, if this is not the case, the original Bernoulli version of the Mixture Model Adjustment appears to offer more comparable adjusted network statistics.  Further, note that using the original Bernoulli version of the Mixture Model Adjustment for networks generated from the mean degree preserving version (and vice versa) appears to be a poor choice.  Thus, in choosing a parametric model to implement in the Mixture Model Adjustment, it is very important to examine (both empirically and theoretically) whether density or average degree is presevered as network size grows.

Surprisingly, when we increase the complexity and flexibility of the reference model in the Mixture Model Adjustment algorithm to the Hierarchical Bernoulli model, we do not see much improvement and, in fact, the adjusted statistics look slightly less comparable across network sizes in many cases.  Although the centrality measures look relatively comparable across network sizes, the topological network statistics' distributions still show evidence of some dependence on $n$, particularly when calculated on non-Bernoulli graphs.

There are two plausible explanations for this behavior.  First, perhaps this model is simply too complex to fit and simulate from in an automated way, as is required by the Mixture Model Adjustment.  That is, perhaps the adjusted statistics would benefit from more model fine-tuning, such as trying different values of $K_{max}$ or respecifying prior parameter values.  Secondly, while the Hierarchical Bernoulli model is logically appealing and perhaps more suited to real-world network data than the simple Bernoulli model, it may be a poor model for the \textit{simulated} datasets examined here.  However, its mediocre performance for network statistics calculated on networks simulated from the Hierarchical Bernoulli distribution itself is curious.


\begin{figure*}[t]
\caption{Boxplots of Network Statistics from Two Hierarchical Markov ERGMs}
\centering
  \centering
  \includegraphics[width=.95\textwidth]{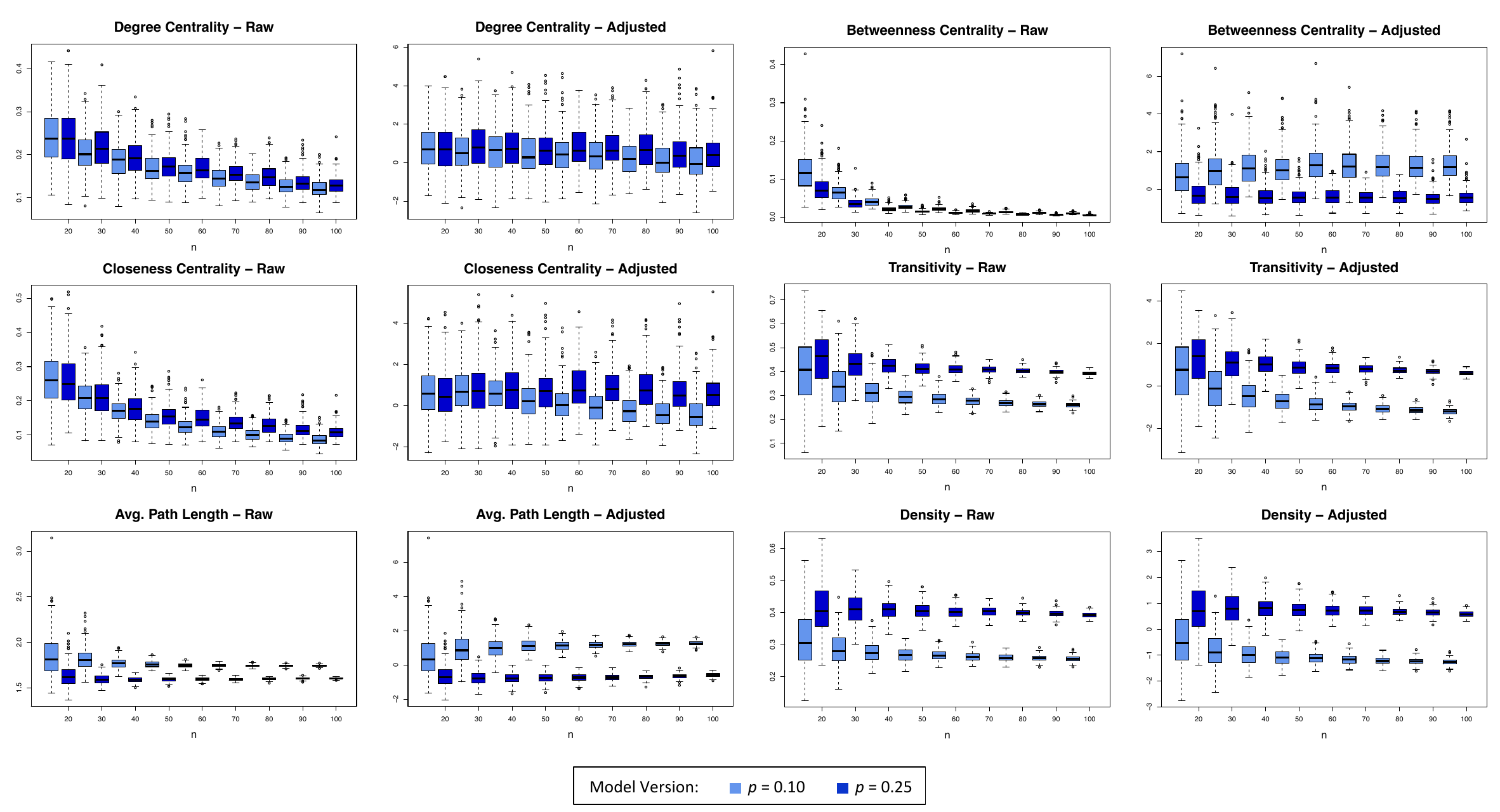}
\vskip1ex
\caption*{The rows indicate the common network statistics, while the left column gives the unadjusted statistic and the right column gives those adjusted by a Bernoulli mixture model.  In each plot, the x-axis gives the network sizes. The color indicates the two versions of the model, where the probability of connections between neighborhoods is given in the legend on the right.}
\label{fig:Boxplot}
\end{figure*}

\section{Simulation Study: Feature Detection} \label{sec:Sim2}
We have demonstrated that the mixture model adjustment can be used to produce adjusted network statistics that are amenable to comparison across networks of different size.  In this section, we design and implement a small simulation study to confirm that the adjusted network statistics can still be used in much the same way as the original, raw network statistics themselves:  to indentify relevant structural features of observed networks.

We simulate networks from two slightly different models, use the Mixture Model Adjustment to produce adjusted versions of our common network statistics, and examine these statistics to check that our adjusted statistics are able to detect the difference between the networks generated from different models, at least to the extent that the unadjusted statistics are capable of detecting the difference.  We use our most realistic generative model, the Hierarchical Markov ERGM, under the same settings described in Section \ref{sec:models} but change $\theta_{btw}$ to create a second version of this model.  More specifically, we consider two versions of the Hierarchical Markov ERGM here:  one where the probability of connections between neighborhoods is $0.10$ and one where this probability is $0.25$.  We simulate $N_n=250$ replicates from these two Hierarchical Markov ERGMs with network sizes $n=20, 30, ... 100$, for a total of $N=2250$ simulated networks per generative model.  Given its superior performance in the simulation studies in the previous section, we use the Bernoulli model to produce the mixture reference distribution and obtain adjusted network statistic values.

We provide boxplots of the results in Figure \ref{fig:Boxplot}.  We are primarily concerned with whether perceivable differences between the two generative models (with $\theta_{btw}=0.10$ and $\theta_{btw}=0.25$) which are evident in the boxplots of the raw, unadjusted statistics are either diminished, preserved, or amplified by the mixture model adjustment.  While the empirical distribution of degree centrality appears to remain roughly the same under either model, we see differences in the empirical distributions of the other unadjusted/adjusted network statistics.  Not surprisingly, in most cases, this difference is harder to distinguish among small networks but increasingly easier to distinguish as the network grows in size.  Note that there is a difference between the empirical distributions of unadjusted transitivity and density statistics across the two generative models, and this distinction is preserved in the adjusted versions of these statistics.  For the remaining network statistics - betweenness centrality, closeness centrality, and average path length - the difference between the empirical distributions of the unadjusted statistics across the two generative models is actually amplified and made more noticeable by the adjustment.

Although this simulation study is rather limited in scope, we expect to see similar results under different settings.  Once the network statistics' dependence on network size is removed, it is not surprising that we might be able to more easily pick up ``true'' differences between networks under comparison.  Not only does the mixture model adjustment allow for comparison of common network statistics across networks of different size, but it also improves the ease of the comparative analysis itself, since differences among adjusted statistics are often more pronounced.


\section{Application: L.A.FANS Ecological Networks} \label{sec:LAFANS}

Finally, we apply the mixture model adjustment to neighborhood network data from the Los Angeles Family and Neighborhood Survey (L.A.FANS).  Conducted in the early 2000s, L.A.FANS is a two-wave survey of roughly 3,000 families in 65 census tracts in Los Angeles County, California \citep{sastry_ghosh-dastidar_adams_2006}.  Here, we use activity data collected during the first wave (2000 and 2001).  Survey participants were asked to report the locations where they conduct a few specific routine activities, such as where they go grocery shopping, where they go to the doctor and where they work.

Following earlier analysis with this dataset, we assign the reported activity locations to census block groups \citep{browning_calder_krivo_etal_2014}.  Naturally, this data can be represented as a two-mode ecological network within each census tract, with the first mode being people who reside in a census tract and the second mode being the census block groups in which activity locations that these people reported visiting are physically located.  We focus only on the one-mode projection of these networks, where people in a tract are connected if they report visiting activity locations that occur in the same census block group.

\begin{figure*}[t]
\caption{L.A.FANS Network Structure}
\centering
\includegraphics[width=.16\textwidth]{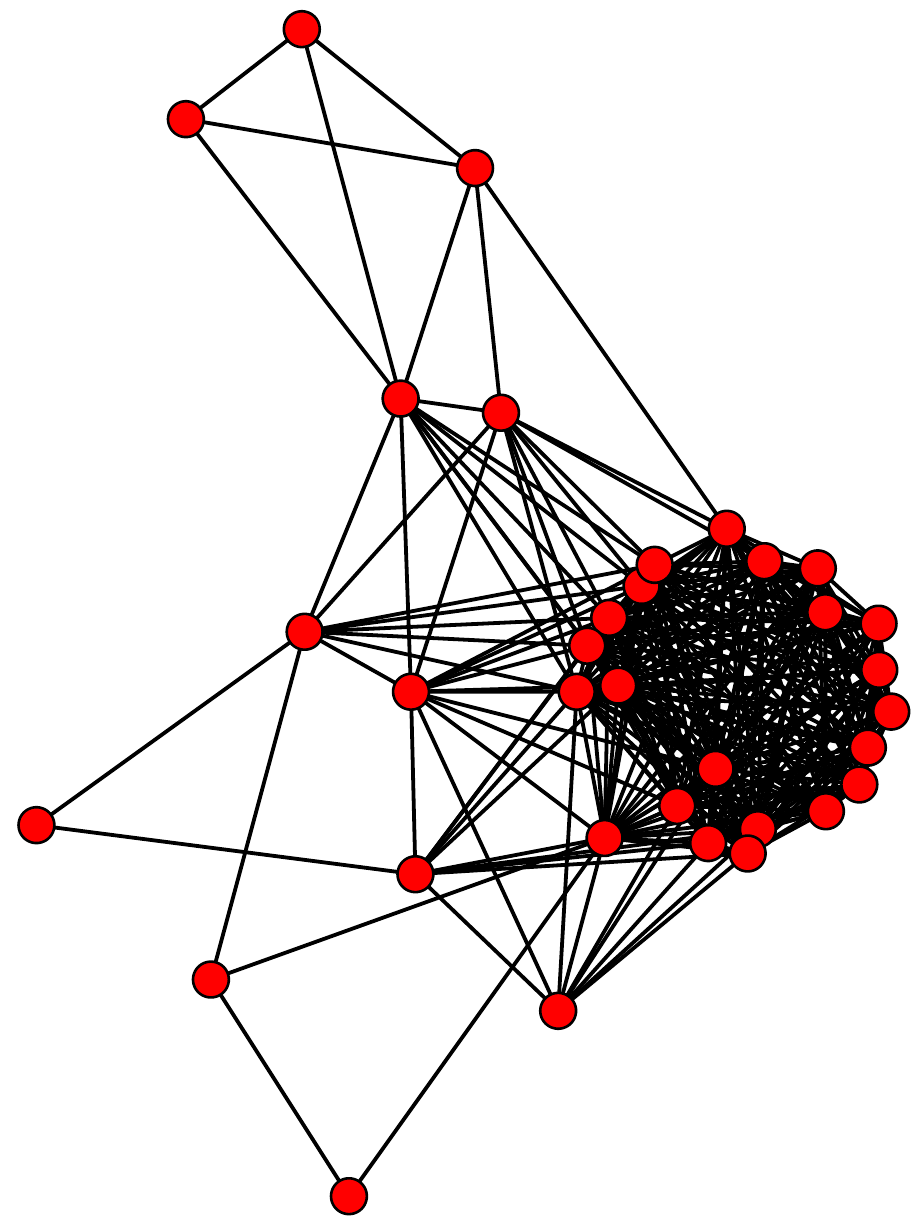}
\includegraphics[width=.27\textwidth]{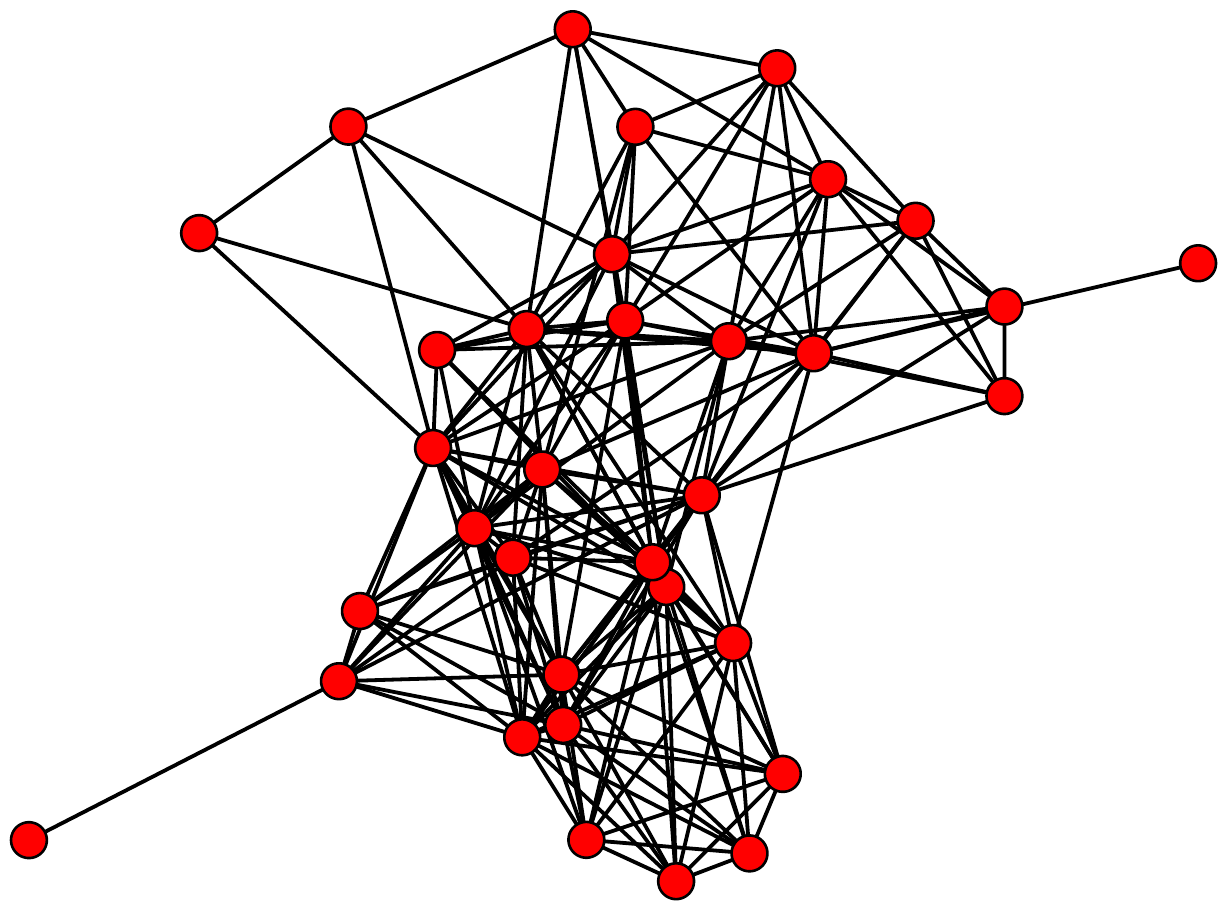}
\caption*{\\The above networks are one-mode projections of activity pattern data from L.A.FANS.  For both of these networks, $n=34$}.
\label{fig:LAFANS_intro}
\end{figure*}

Note that we observe $N=65$ networks, where the nodes in each network represent the survey participants from a particular census tract.  The network sizes range from $n=27$ to $n=54$ people.  Visually, we notice two structural patterns in these data.  We give examples of these two patterns in Figure \ref{fig:LAFANS_intro}\footnote{The identity of the sampled census tracts in L.A.FANS is withheld to protect respondents' identity.}.  For the most part, these 65 observed networks look like either of the two networks in this figure.  Either the network consists of a single group of very highly connected nodes and a few nodes with only a few connections (i.e. it exhibites some type of core-periphery structure) or is generally less dense, with multiple groups of more highly connected nodes.

\begin{figure}[h]
\caption{Density and Mean Degree in L.A.FANS Networks}
\centering
\includegraphics[width=.75\textwidth]{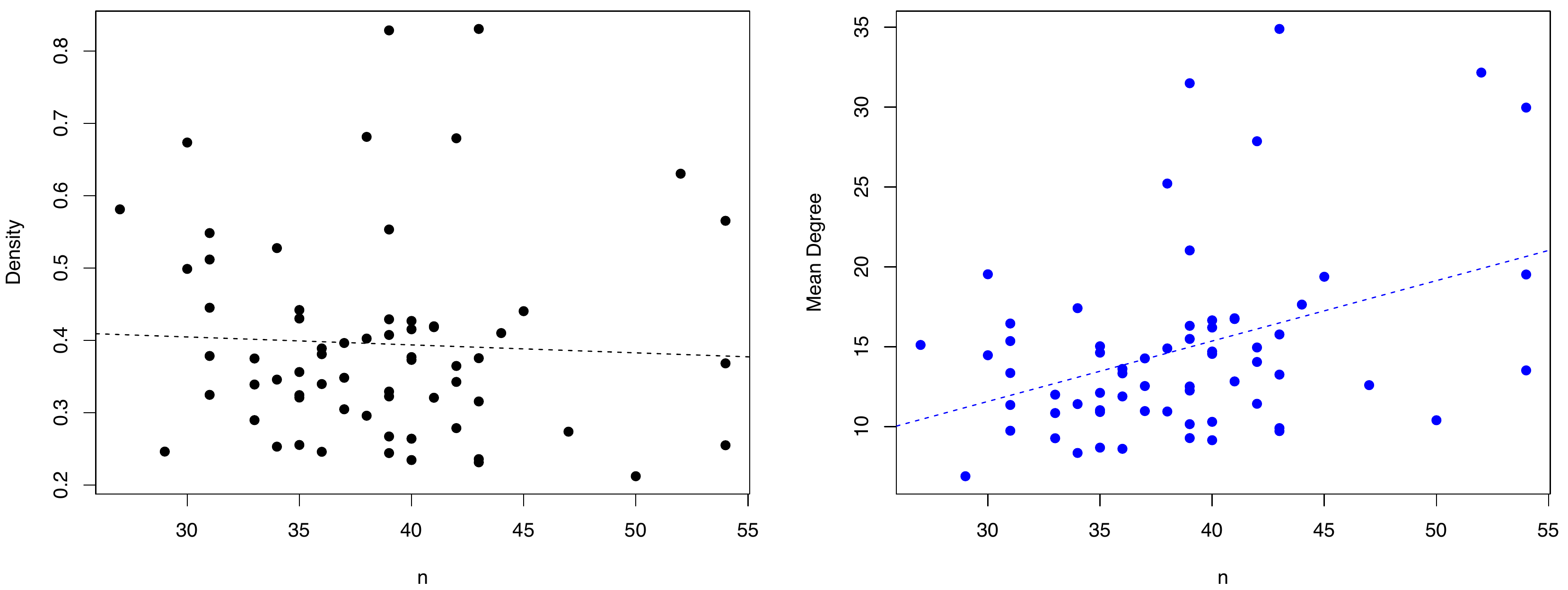}
\caption*{Density (left) and mean degree (right) are plotted against network size for the $N=65$ observed L.A.FANS census tracts.  The dotted lines represent fitted simple linear regressions.}
\label{fig:LAFANS_dens}
\end{figure}

We analyze proporties of these networks by mirroring the structure of this paper thus far:  first, we examine the raw network statistics themselves; secondly, we standardize relative to a fully parameterized reference distribution, \ER graphs; finally, we apply the Mixture Model Adjustment where we use the Bernoulli model, the mean degree preserving Bernoulli model and the Hierarchical Bernoulli model to produce the mixture reference distributions.  We use all of the same algorithm settings as described before.  As was observed in the simulation study, we expect the performance of the Hierarchical Bernoulli mixture to be poor relative to its computational requirements.  Further, recall that the simulation study results indicated that using the mean degree preserving version of the Bernoulli model would be superior to the original version if mean degree (rather than density) is preserved as network size increases.  We consider a simple plot of density and mean degree against network size in Figure \ref{fig:LAFANS_dens}.  Since the density of our observed networks appears to be preserved as network size changes, we expect the mixture model adjustment using Bernoulli model components to be superior to the version which includes \citet{krivitsky_handcock_morris_2011}'s offset term.

Unlike in the simulation studies, here, we no longer have much replication across network size, and so we turn to new metrics for evaluating how comparable the network statistics are across different network sizes.  In Figure 9 in the Supplementary Materials we provide simple plots of the network statistics by network size, $n$.  If the network statistics are comparable across different network sizes, we would not expect to see dependence between network statistics and $n$.  Or rather, we would expect to see the correlation between the raw statistics and $n$ be reduced by our adjustment methods.  We provide a table of such correlations in Table \ref{tab:LAFANScorr}.

\begin{table}[h]
\caption{L.A.FANS Data:  Correlation between Network Statistics and Network Size}
\footnotesize
\centering
\begin{tabular}{c|l|rrrrr}
\multicolumn{1}{l}{} && \multicolumn{5}{c}{\textsc{Adjustment Model}}\\
  \cline{3-7}
\multicolumn{1}{l}{} && \multirow{2}{*}{Unadjusted} & \multirow{2}{*}{\ER} & \multirow{2}{*}{Bernoulli} & Mean Degree & Hierarchical \\  
\multicolumn{1}{l}{} && & & & Bernoulli & Bernoulli \\
 \hline
  \hline
\parbox[t]{2mm}{\multirow{6}{*}{\rotatebox[origin=c]{90}{ \textsc{Statistic} }}} & Deg. Cent. & 0.8228 & 0.3005 & 0.3026 & 0.3509 & 0.1877 \\ 
 & Betw. Cent. & 0.6598 & 0.4670 & 0.4290 & 0.1351 & 0.3851 \\ 
 & Clos. Cent. & 0.2228 & 0.0745 & 0.1060 & 0.1299& 0.0417 \\ 
 & Transitivity & -0.0981 & 0.2543 & -0.0926 & 0.2286 & -0.1000 \\ 
 & Avg. Path & 0.0081 & 0.1224 & 0.0651 & -0.1659 & 0.0236 \\ 
&  Density & -0.0473 & -0.1298 & -0.0473 & 0.1878 & -0.0522 \\ 
\end{tabular}
\label{tab:LAFANScorr}
\end{table}

Although the plotting scale obviously differs across $n$, the adjustment methods do not seem to have a strong effect on the overall shape or pattern that we observe in any of these plots, with the exception of degree centrality.  For the unadjusted degree centrality measures, we see a strong relationship with network size.  However, applying any kind of adjustment method seems to ammeliorate the issue.  For all other network statistics, we do not observe such an obvious improvement.

However, when we examine the correlations in Table \ref{tab:LAFANScorr}, we see that applying some type of an adjustment method does reduce the correlation between network size and many of the network statistics.  This holds for all of the centralization measures, but is violated for average path length as well as \ER adjusted transitivity and density.  Recall that in our simulation study, the \ER adjustment performed rather poorly for topological measures, so it is not surprising that it appears to perform poorly for these same measures in this application as well.  Further, although the correlation between average path length and network size is not reduced by the adjustment methods, it is very near zero to begin with and remains relatively near zero under both the Bernoulli and Hierarchical Bernoulli versions of the Mixture Model Adjustment.  

Given these results and the conclusions from our simulation study, we recommend using network statistics adjusted by the Bernoulli version of the Mixture Model Adjustment in any further analysis involving these network statistics.  Note that all versions of the Mixture Model Adjustment performed comparably, but that the Bernoulli version is much less computationally expensive (see Table 5 in the Supplementary Materials) than the Hierarchical Bernoulli version, and the assumptions of the Bernoulli model (constant density across network size) appear to be a good match for our observed data.

Moving forward with our analysis, we now use the adjusted network statistics to compare across networks of different size.  Note that although the adjusted network statistics are z-scores, they should not be interpretted as conveying any type of statistical significance.  Rather, they are measures of \textit{relative} comparison (much like PCA component scores), that pick up differences among the networks being compared.  Finally, we will briefly examine ways in which using these adjusted values can impact the types of conclusions we might draw about these data.

\begin{figure*}[t]
\caption{Degree Centrality in the L.A.FANS Neighborhood Data}
\centering
\includegraphics[width=.85\textwidth]{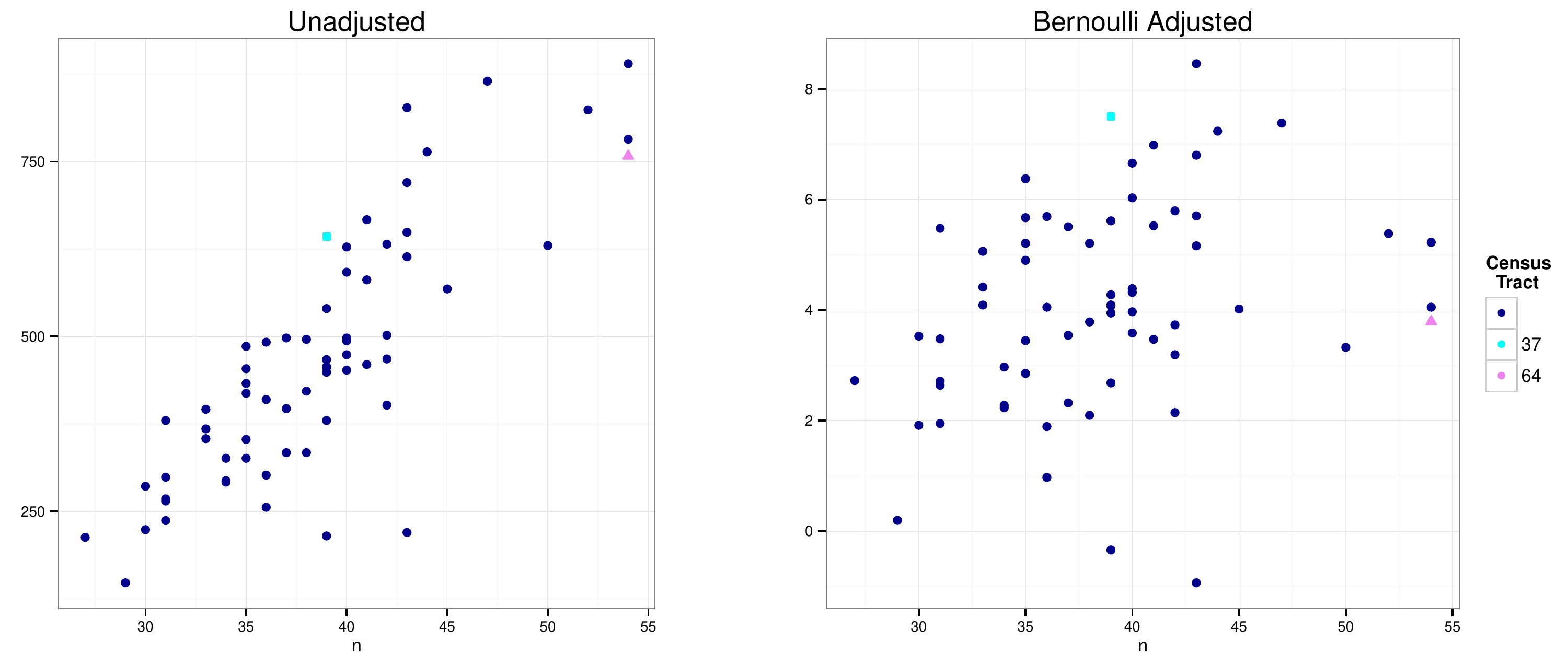}
\label{fig:LAFANS_Deg}
\caption*{The horizontal axis denotes network size, $n$, while the vertical axis gives the statistic value.}
\end{figure*}

For example, if we examine the raw degree centralization values across our group of observed networks, we notice that the network for census tract \#64 has a larger value than that for census tract \#37 (see Figure \ref{fig:LAFANS_Deg}).  However, it turns out that this observation is entirely driven by network size.  Once we perform the Bernoulli version of the Mixture Model Adjustment, we see the opposite:  the network for census tract \#37 has a much larger degree centrality value than we would expect for networks of that size while census tract \#64's is closer to what we might expect for its size.  And recall, that in this type of statement, we are comparing our observed networks to simulated networks of the same size which mimic the type of structure that we observe in our dataset (here, the only structural effect we are mimicking is density, since we are using the Bernoulli version of the Mixture Model Adjustment).  Thus, we might conclude that, relative to the type of network data we observed, census tract \#37 has higher degree centrality than census tract \#64, net of network size effects.  These types of rank changes are common throughout the various network statistics examined here.

In summary, using unadjusted or poorly adjusted (i.e. via the \ER adjustment) network statistics in a comparative analysis of network data can lead to very different conclusions than if one were to utilize our proposed adjusted statistics.  In fact, we argue that these different conclusions could in fact be dangerously misleading because the statistics are not properly adjusted for network size.  Thus, we encourage using a Bernoulli version (perhaps including \citet{krivitsky_handcock_morris_2011}'s offset term) of the Mixture Model Adjustment to first adjust observed network statistics for size effects and then proceed with any desired analysis of these common network statistics per usual (e.g. as independent variables in a regression analysis).  The decision to include \citet{krivitsky_handcock_morris_2011}'s offset term could be guided by theoretical considerations of the expected behavior of density and mean degree for the particular type of network data under consideration, as network size increases.  This decision should also be influenced by any patterns in the observed data, as was considered here.  That is, simply plotting density and mean degree against network size for the observed networks should help inform whether or not an offset term is helpful.  


\section{Discussion}
We have shown that network statistics are not amenable to direct comparison across networks of different size, even when the networks under consideration are generated from the same model.  Although we can not make this statement absolutely, we have demonstrated this phenomenon across a range of popular and reasonable generative models for network data as well as across a variety of popular network statistics.

We previewed the advantages of the Mixture Model Adjustment and have shown that in many cases the adjusted statistics are more comparable across network sizes and are still capable of describing interesting features of networks.  This provides researchers with a reliable way of making comparisons across networks of different sizes, without the difficulty of specifying a fully parameterized reference distribution a priori.  The resulting adjusted network statistics can be interpretted as measures of relative comparison among the observed networks, similar to PCA component scores.

Due to its superior performance and low computational cost, we have recommended use of the Bernoulli model (perhaps including \citet{krivitsky_handcock_morris_2011}'s offset term) for the components of the mixture reference distribution and provided a simple approach for deciding whether or not to include the offset term.  Perhaps implementing network models other than those examined here as mixture components in this adjustment will offer even greater improvements, or perhaps certain network statistics are more amenable to adjustments utilizing a particular class of network models.  We leave this topic for future investigation.

The strong performance of the Bernoulli models is not surprising.  \citet{anderson_butts_carley_1999, faust_2006} and \citet{vanwijk_stam_daffertshofer_2010} all document a relationship between network density (or, closely related, average degree, as examined by \citeauthor{anderson_butts_carley_1999, vanwijk_stam_daffertshofer_2010}) and a variety of graph-level network statistics.  Thus, it is not surprising that a method which properly adjusts for network density would render the network statistics more comparable across network sizes.

Of course, the range of network statistics and generative network models examined here could be expanded as well.  We have considered only statistical generative network models rather than mathematical models, such as the conditional uniform graphs examined by \citet{anderson_butts_carley_1999}.  Our intent was to create a mixture reference distribution which represents the type of structure that we \textit{expect} to see in, say, another realization of our set of observed networks, rather than viewing the observed network structure as some \textit{fixed} attribute and adjusting for it, i.e. rather than conditioning on \textit{observed} network attributes as is done in CUG models.

Further examination of the performance of the Mixture Model Adjustment itself is also needed.  In particular, a consideration of the effect of $N_M$, the number of mixture components, could prove to be a worthwhile investigation.  Other measures of distributional comparison could be used, rather than the simple z-score which performs poorly for nonnormal distributions.  Further, perhaps mixture weights could be incorporated within the Mixture Model Adjustment, proportional to some measure of model fit, to better adjust for observed variability in the network dataset.


\section*{Acknowledgements}
Support for this work was provided by grants from the National Science Foundation (NSF DMS-1209161), the National Institute of Health (NIH R01DA032371), the William T. Grant Foundation, and The Ohio State University Institute for Population Research (NIH P2CHD058484).


\def\url#1{}
\bibliography{compare}


\end{document}